\documentclass[seceq]{ptptex}

\usepackage[dvips]{graphicx}
\usepackage{bm}

\newtheorem{theorem}{Theorem}[section]
\newtheorem{lemma}{Lemma}[section]
\newtheorem{prop}{Proposition}[section]

\newcommand{\hook}
{\raisebox{-0.35ex}{\makebox[0.6em][r]{\scriptsize $-$}}
\hspace{-0.15em}\raisebox{0.25ex}{\makebox[0.4em][l]{\tiny $|$}}}

\title{%        %You can use \\ for explicit line-break.
Hidden Symmetry and Exact Solutions in Einstein Gravity}

\author{%       %Use \scshape for the family name.
Yukinori \textsc{Yasui}$^a$ and Tsuyoshi \textsc{Houri}$^b$%
}

\inst{%     %Affiliation, neglected when [addenda] or [errata].
$^a$Department of Mathematics and Physics, Graduate School of Science, \\
Osaka City University, 3-3-138 Sugimoto, Sumiyoshi, Osaka 558-8585, JAPAN
$^b$Osaka City University Advanced Mathematical Institute (OCAMI), \\
3-3-138 Sugimoto, Sumiyoshi, Osaka 558-8585, JAPAN \\[10pt]

}

\abst{%         %This abstract is neglected when [addenda] or [errata].
Conformal Killing-Yano tensors are introduced as a generalization of Killing vectors.
They describe symmetries of higher-dimensional rotating black holes.
In particular, a rank-2 closed conformal Killing-Yano tensor 
generates the tower of both hidden symmetries and isometries.
We review a classification of higher-dimensional spacetimes
admitting such a tensor, and
present exact solutions to the Einstein equations for these spacetimes.
}

\begin{document}
\maketitle

\vspace{-7.5cm}
\hfill{OCU-PHYS 350}
\vspace{7.5cm}

\tableofcontents

\section{Introduction}
Symmetries of spacetimes play an important role 
in the study of exact solutions of Einstein equations.
Killing vectors describe isometries on the spacetimes, which are the most fundamental continuous symmetries.
If spacetimes have enough isometries, 
one can expect that the Einstein equations become a system of
algebraic or ordinary differential equations 
and then one may find the general solutions comparatively easily.
Recently, inspired by the supergravity theories and string theories 
a large number of higher-dimensional rotating black hole solutions 
have been found \cite{Hawking:1999,Gibbons:2004a,Gibbons:2005,Chen:2006b,Chen:2007,Myers:1986}.
The most general known vacuum solution is the higher-dimensional Kerr-NUT-(A)dS metric \cite{Chen:2006b}.
Although these exact solutions have been constructed, 
an organizing  principle is still lacking, and also the isometries
are not usually enough to characterize the solutions.
What would be a generalization of the Killing vector 
which is effective in higher-dimensional spacetimes?

We begin with a brief review of the four-dimensional Kerr geometry.
Arguably, ``hidden symmetries'' of the Kerr spacetime would lead us to generalizations of Killing vectors.
One of the most remarkable properties of the Kerr spacetime
is separation of variables in the equations for 
a free particle, scalar, Dirac and Maxwell fields and gravitational perturbations,
\cite{Carter:1968a,Carter:1968b,Teukolsky:1972,Teukolsky:1973,Unruh:1973,Chandrasekhar:1976,Page:1976}
while not enough isometries are present. 
The existence of a Killing-Yano tensor explains 
these integrabilities within the geometric framework; 
besides isometries the Kerr spacetime possesses
hidden symmetries generated by the Killing-Yano tensor.
\cite{Walker:1970,Floyd:1973,Penrose:1973,Hughston:1973}

In the 1950s and 1960s Killing-Yano tensors and conformal Killing-Yano tensors, 
which are generalizations of Killing vectors and conformal Killing vectors respectively, 
were investigated by Japanese geometricians
\cite{Bochner:1948,Yano:1952,Yano:1953,Kashiwada:1968,Tachibana:1969}. 
In spite of interest to continue for a long time, 
much less is known about these tensors. 
This article attempts to move this situation forward.
We will see that conformal Killing-Yano (CKY) tensors 
are successfully applied to describe symmetries of higher-dimensional black hole spacetimes.
In particular, a rank-2 closed CKY tensor
generates the tower of both hidden symmetries and isometries\cite{Page:2007,Krtous:2007a,Krtous:2007b}. 
We present a complete classification of
higher-dimensional spacetimes admitting such a tensor\cite{Houri:2009}, 
and further obtain exact solutions to the Einstein equations\cite{Houri:2008b}. 

As an interesting case one can consider a special CKY tensor with maximal order, 
which we call a principal CKY tensor. 
It is shown that the Kerr-NUT-(A)dS spacetime is the only Einstein space 
admitting a principal CKY tensor\cite{Houri:2007,Krtous:2008}. 
For general (possibly degenerate) rank-$2$ closed CKY tensor the geometry is much richer, 
and the metrics are written as
``Kaluza-Klein metrics'' on the bundle over K\"ahler manifolds whose fibers are Kerr-NUT-(A)dS spacetimes. 
It is remarkable that a so-called Wick rotation transforms these metrics 
into complete (positive definite) Einstein metrics without singularities
\cite{Page:1979,Hashimoto:2005,Gibbons:2005}.
We also briefly discuss an extension of this classification\cite{Kubiznak:2009a,Houri:2010b}, 
where a skew-symmetric torsion is introduced. 
The spacetimes with torsion naturally occur in supergravity theories and 
string theories, 
and then the torsion may be identified with a $3$-form field strength.
\cite{Strominger:1986}

The paper is organized as follows. After reviewing the four-dimensional Kerr geometry in section 2, 
we introduce CKY tensors in section 3, and their basic properties are discussed.
In section 4 we describe a higher-dimensional generalization of the Kerr geometry.
Section 5 is a central part of this paper. 
A complete classification of spacetimes admitting a CKY tensor 
is given by theorems \ref{uniqueness}--\ref{theorem:5.4}.
Finally, section 6 involves two components of independent interest. 
In the first part we review a Killing-Yano symmetry in the presence of torsion. 
In the second part, as an application of theorems \ref{theorem:5.3} and \ref{theorem:5.4}, 
we present existence theorems \ref{theorem:6.1} and \ref{theorem:6.2} of Einstein metrics on compact manifolds.
Section 7 is devoted to summary.

Many of results presented here are already available in the literature. 
However, we collect them in a systematic way with a particular emphasis on exact solutions and symmetries.

\subsection*{Notations and Conventions}
In this chapter we use a mixture of invariant and tensorial notation. 
The tensors are denoted in boldface, as $\bm{\xi}$, $\bm{g}$, $\cdots$, 
and their components are in normal letters.
Indices $a,b,\cdots$ are used for abstract index and
indices $M,N,\cdots$ for components in a certain local coordinates $q^M$
on $D$-dimensional spacetime $({\cal M},\bm{g})$.
Especially, a differential $p$-form (a rank-$p$ antisymmetric tensor) $\bm{k}$ denotes
\begin{align}
\bm{k} = \frac{1}{p!}\,k_{M_1\cdots M_p}\,dq^{M_1}\wedge\cdots\wedge dq^{M_p} ~,
\end{align}
where $\wedge$ stands for the Wedge product.
As the differential operator, 
we use the exterior derivative $d$, co-exterior deivetive $\delta$ and Hodge star $*$ 
mapping a $p$-form $\bm{k}$, respectively, 
into a $(p+1)$-form $d\bm{k}$, a $(p-1)$-form $\delta\bm{k}$ and a $(D-p)$-form $*\bm{k}$ as
\begin{align}
& (dk)_{a_1a_2\cdots a_{p+1}} = (p+1)\,\nabla_{[a_1}k_{a_2\cdots a_{p+1}]} ~, \nonumber\\
& (\delta k)_{a_1\cdots a_{p-1}} = -\nabla^bk_{ba_1\cdots a_{p-1}} ~, \nonumber\\
& (*k)_{a_1\cdots a_{D-p}}=\frac{1}{p!}\,\varepsilon_{a_1\cdots a_{D-p}}{}^{b_1\cdots b_p}k_{b_1\cdots b_p} ~,
\end{align}
where $\varepsilon_{a_1\cdots a_D}$ is the $D$-dimensional Levi-Civita tensor.

\section{``Hidden Symmetry'' of the Kerr Black Hole}

\subsection{Symmetries in the Kerr spacetime}
In 1963, Kerr \cite{Kerr:1963} discovered  a stationary and axially symmetric solution,
which is describing a rotating black hole in a vacuum.
The Kerr metric is written in the Boyer-Lindquist's coordinates as
\begin{align}
ds^2
=& -\frac{\triangle}{\Sigma}\Big(dt-a\sin^2\theta d\phi\Big)^2 \nonumber\\
 & +\frac{\sin^2\theta}{\Sigma}\Big(a\,dt-(r^2+a^2)d\phi\Big)^2
   +\frac{\Sigma}{\triangle}\,dr^2+\Sigma\,d\theta^2 ~, \label{Kerr metric}
\end{align}
where
\begin{align}
\Sigma=r^2+a^2\cos^2\theta ~,~~~
\triangle = r^2-2Mr+a^2 ~.
\end{align}
This metric admits two isometries, $\bm{\partial}_t$ and $\bm{\partial}_\phi$, 
which corresponds, respectively, to the time translation and the rotation. 
The parameters $M$ and $a$ are responsible for
the mass $M$ and the angular momentum $J=Ma$ of the black hole.
When the black hole stops rotating, i.e., $a=0$,
a static and spherically symmetric solution, 
Schwarzschild metric \cite{Schwarzschild:1916}, is obtained.

It is convenient to introduce an orthonormal basis $\{\bm{e}^\mu\}$ $(\mu=0,1,2,3)$ 
from the view point of hidden symmetries.
For the Kerr metric (\ref{Kerr metric}), we choose it as
\begin{align}
&  \bm{e}^0 = \frac{\sqrt{\triangle}}{\sqrt{\Sigma}}\Big(dt-a\sin^2\theta d\phi\Big) ~,
&& \bm{e}^1 = \frac{\sqrt{\Sigma}}{\sqrt{\triangle}}\,dr ~, \nonumber\\
&  \bm{e}^2 = \frac{\sin\theta}{\sqrt{\Sigma}}\Big(a\,dt-(r^2+a^2)d\phi\Big) ~, \label{4DOB}
&& \bm{e}^3 = \sqrt{\Sigma}\,d\theta ~,
\end{align}
in which the metric is written as
\begin{align}
ds^2 = -\bm{e}^0\bm{e}^0+\bm{e}^1\bm{e}^1+\bm{e}^2\bm{e}^2+\bm{e}^3\bm{e}^3 ~.
\end{align}
The inverse basis $\{\bm{e}_\mu\}$ is given by
\begin{align}
& \bm{e}_0 = \frac{1}{\sqrt{\Sigma\triangle}}\Big((r^2+a^2)\bm{\partial}_t+a\,\bm{\partial}_\phi\Big) ~,
&&\bm{e}_1 = \frac{\sqrt{\triangle}}{\sqrt{\Sigma}}\,\bm{\partial}_r ~, \nonumber\\
& \bm{e}_2 = -\frac{1}{\sqrt{\Sigma}\sin\theta}\Big(a\sin^2\theta\,\bm{\partial}_t+\bm{\partial}_\phi\Big) ~, 
&&\bm{e}_3 = \frac{1}{\sqrt{\Sigma}}\,\bm{\partial}_\theta ~. \label{4DIB}
\end{align}
From the first structure equation
\begin{align}
d\bm{e}^\mu+\bm{\omega}^\mu{}_{\nu}\wedge\bm{e}^\nu=0 
\end{align}
and $\omega_{\mu\nu}=-\omega_{\nu\mu}$, we obtain the connection 1-forms
\begin{align}
&\bm{\omega}_{01} = -A\bm{e}^0-B\bm{e}^2 ~,
&&\bm{\omega}_{02} = -B\bm{e}^1+C\bm{e}^3 ~,
&&\bm{\omega}_{03} = -D\bm{e}^0-C\bm{e}^2 ~, \nonumber\\
&\bm{\omega}_{12} = B\bm{e}^0-E\bm{e}^2 ~,
&&\bm{\omega}_{13} = D\bm{e}^1-E\bm{e}^3 ~,
&&\bm{\omega}_{23} = -C\bm{e}^0-F\bm{e}^2 ~, \label{4DCN}
\end{align}
where
\begin{align}
&A=\frac{d}{dr}\Big(\frac{\sqrt{\triangle}}{\sqrt{\Sigma}}\Big) ~,
&&B=\frac{ar\sin\theta}{\Sigma\sqrt{\Sigma}} ~,
&&C=\frac{a\cos\theta\sqrt{\triangle}}{\Sigma\sqrt{\Sigma}} ~, \nonumber\\
&D=-\frac{a^2\sin\theta\cos\theta}{\Sigma\sqrt{\Sigma}} ~,
&&E=\frac{r\sqrt{\triangle}}{\Sigma\sqrt{\Sigma}} ~,
&&F=-\frac{1}{\sin\theta}\frac{d}{d\theta}\Big(\frac{\sin\theta}{\sqrt{\Sigma}}\Big) ~.
\end{align}

\subsubsection{Separation of variables in the field equations}
In 1968, it was demonstrated by Carter \cite{Carter:1968a, Carter:1968b} that
in a class of solutions of Einstein-Maxwell equations including the Kerr spacetime,
both of the Hamilton-Jacobi and the Schr\"odinger equation (as the scalar field equation) 
can be solved by separation of variables.
This means that there exists a forth integral of geodesic motion known as Carter's constant,
apart from integrals associated with two Killing vectors and the Hamiltonian.
The geodesic motion of a particle is governed by the Hamilton-Jacobi equation for geodesics
\begin{align}
\partial_\lambda S+g^{ab}\partial_aS\,\partial_bS = 0 ~.
\end{align}
For the Kerr metric (\ref{Kerr metric}), this equation takes the following explicit form
\begin{align}
\frac{\partial S}{\partial \lambda}
-\frac{1}{\Sigma\triangle}\Big((r^2+a^2)\partial_tS+a\,\partial_\phi S\Big)^2
+\frac{\triangle}{\Sigma}(\partial_rS)^2& \nonumber\\
+\frac{1}{\Sigma\sin^2\theta}(a\sin^2\theta\,\partial_tS+\partial_\phi S)^2
+\frac{1}{\Sigma}(\partial_\theta S)^2&=0 ~.
\end{align}
Then, we find that it allows the additive separation
\begin{align}
S = -\kappa_0\lambda-Et+L\phi+R(r)+\Theta(\theta) ~,
\end{align}
and hence the functions $R(r)$, $\Theta(\theta)$ obey the ordinary differential equations
\begin{align}
\Big(\frac{dR}{dr}\Big)^2 -\frac{W_r^2}{\triangle^2}-\frac{V_r}{\triangle}=0 ~,~~~
\Big(\frac{d\Theta}{d\theta}\Big)^2 +\frac{W_\theta^2}{\sin^2\theta}-V_\theta=0 ~, \label{ode HJ Kerr}
\end{align}
where
\begin{align}
&  W_r=-E(r^2+a^2)+aL ~,
&& V_r=\kappa+\kappa_0 \,r^2 ~, \nonumber\\
&  W_\theta=-aE\sin^2\theta+L ~,
&& V_\theta=-\kappa+\kappa_0\,a^2\cos^2\theta ~. \label{pot HJ Kerr}
\end{align}
As a consequence, we obtain the momentum of the particle $p_a=\partial_aS$
\begin{align}
\bm{p} = -E\,dt+L\,d\phi
      \pm\sqrt{\frac{W_r^2}{\triangle^2}+\frac{V_r}{\triangle}}\,dr
      \pm\sqrt{V_\theta-\frac{W_\theta}{\sin^2\theta}}\,d\theta ~,
\end{align}
where the two $\pm$ signs are independent, parameters $E$ and $L$ are separation constants 
corresponding to the Killing vectors $\bm{\partial}_t$ and $\bm{\partial}_\phi$, 
$\kappa_0$ is the normalization of the momentum.
As explained in the next section, 
a rank-2 irreducible Killing tensor exists in the Kerr spacetime.
The parameter $\kappa$ is interpreted as a separation constant
associated with the Killing tensor (\ref{4DKT}),
which always appears when variables are separated.

Similarly, separation of variables occurs 
in the massive scalar field equation (the massive Klein-Gordon equation),
\begin{align}
\square \Phi - m^2\Phi = 0 ~. \label{4DKG}
\end{align}
Making use of the expression 
$\square\Phi=\displaystyle{\frac{1}{\sqrt{-g}}}\partial_a(\sqrt{-g}g^{ab}\partial_b\Phi)$, 
we find that it allows multiplicative separation
\begin{align}
\Phi = e^{-i\omega t+in\phi}R(r)\Theta(\theta) ~.
\end{align}
The ordinary differential equations for the functions $R(r)$ and $\Theta(\theta)$ are
\begin{align}
& \frac{1}{R}\frac{d}{dr}\Big(\triangle\frac{dR}{dr}\Big)
  +\frac{U_r^2}{\triangle}-m^2r^2-\kappa=0 ~, \\
& \frac{1}{\Theta}\frac{1}{\sin\theta}\frac{d}{d\theta}\Big(\sin\theta\frac{d\Theta}{d\theta}\Big)
  -\frac{U_\theta^2}{\sin^2\theta}-m^2a^2\cos^2\theta-\kappa=0 ~,
\end{align}
where the potential functions are given by
\begin{align}
U_r = an-\omega(r^2+a^2) ~,~~~ U_\theta = n-a\,\omega\sin^2\theta ~. \label{4DPU}
\end{align}
As was expected, $\kappa$ has appeared as a separation constant 
associated with the Killing tensor.

It has been shown that not only scalar field equations can be solved by separation of variables.
In 1972, Teukolsky \cite{Teukolsky:1972} decoupled equations 
for electromagnetic field and gravitational perturbation, 
and separated variables in their resulting master equations.
A year later, it was shown that separation of variables occurs in 
massless neutrino equation by Teukolsky \cite{Teukolsky:1973} and Unruh \cite{Unruh:1973}.
In 1976, it was demonstrated by Chandrasekhar \cite{Chandrasekhar:1976} and Page \cite{Page:1976} 
that the massive Dirac equation is separable.
Let us close this section by seeing separation of variables in the massive Dirac equation
\begin{align}
(\gamma^a\nabla_a+m)\Psi = 0 ~, \label{4DDirac}
\end{align}
which reads
\begin{align}
\Big[\gamma^a\Big(e_a+\frac{1}{4}\gamma^b\gamma^c\omega_{bc}(e_a)\Big)+m\Big]\Psi = 0 ~.
\end{align}
By using the inverse basis (\ref{4DIB}) and the connection 1-forms (\ref{4DCN}),
this equation takes the explicit form,
\begin{align}
& \Bigg[\frac{\gamma^0}{\sqrt{\Sigma\triangle}}\Big((r^2+a^2)\partial_t+a\,\partial_\phi\Big)
  +\gamma^1\Big(E+\frac{A}{2}+\sqrt{\frac{\triangle}{\Sigma}}\,\partial_r\Big)
  -\frac{\gamma^2}{\sqrt{\Sigma}\sin\theta}(a\sin^2\theta\partial_t+\partial_\phi) \nonumber\\
& +\gamma^3\Big(D-\frac{F}{2}+\frac{1}{\sqrt{\Sigma}}\,\partial_\theta\Big)
  +\gamma^{012}\frac{B}{2}+\gamma^{023}\frac{C}{2}+m\Bigg]\Psi = 0 ~. \label{4DDirac2}
\end{align}
We further use the following representation of gamma matrices $\{\gamma^a,\gamma^b\}=2\delta^{ab}$:
\begin{align}
& \gamma^0=
\left(
\begin{array}{cc}
0 &\hspace{-0.1cm}-I \\
I &0 
\end{array}
\right) ~,~~~
\gamma^1=
\left(
\begin{array}{cc}
0 &I \\
I &0 
\end{array}
\right) ~, \nonumber\\
& \gamma^2=
\left(
\begin{array}{cc}
\sigma^2 &0 \\
0 &\hspace{-0.1cm}-\sigma^2 
\end{array}
\right) ~,~~~
\gamma^3=
\left(
\begin{array}{cc}
\sigma^1 &0 \\
0 &\hspace{-0.1cm}-\sigma^1 
\end{array}
\right) ~,
\end{align}
where $I$ is the $2\times 2$ identity matrix and $\sigma^i$ are Pauli's matrices.
Separation of the Dirac equation can be achieved with the ansatz
\begin{align}
\Psi = e^{-i\omega t+in\phi}
\left(
\begin{array}{c}
(r+ia\cos\theta)^{-1/2}R_+\Theta_+ \\
(r-ia\cos\theta)^{-1/2}R_+\Theta_- \\
(r-ia\cos\theta)^{-1/2}R_-\Theta_+ \\
(r+ia\cos\theta)^{-1/2}R_-\Theta_- \\
\end{array}
\right) 
\end{align}
with functions $R_\pm=R_\pm(r)$ and $\Theta_\pm=\Theta_\pm(\theta)$.
Inserting this ansatz in (\ref{4DDirac2}), we obtain eight equations with four separation constants.
The consistency of these equations implies that only one of the separation constant is independent, 
we denote it by $\kappa$.
In the end, we obtain the following four coupled first order ordinary differential equations 
for $R_\pm$ and $\Theta_\pm$:
\begin{align}
& \frac{dR_\pm}{dr}+R_\pm \frac{\partial_r \triangle \pm 4iU_r}{\triangle}
  +R_\mp\frac{mr\mp\kappa}{\sqrt{\triangle}}=0 ~, \\
& \frac{d\Theta_\pm}{d\theta}+\Theta_\pm \frac{\cos\theta\pm 2U_\theta}{2\sin\theta}
  +\Theta_\mp(\pm ima\cos\theta-\kappa)=0 ~,
\end{align}
where $U_r$ and $U_\theta$ are given by (\ref{4DPU}).

\subsubsection{Hidden symmetries of the Kerr spacetime}
We have seen that separation of variables occurs 
in the various field equations of the Kerr spacetime.
Meanwhile there were some progress on the hidden symmetries of the Kerr spacetime.
In 1970, in the transparent fashion rather than Carter, 
it was proved that 
in the Kerr spacetime the Hamilton-Jacobi equation can be integrated.
Walker and Penrose \cite{Walker:1970} pointed out that
the Kerr spacetime admits a rank-2 irreducible Killing tensor \cite{Stackel:1895} obeying
\begin{align}
K_{ab}=K_{(ab)} ~,~~~ \nabla_{(c}K_{ab)}=0 ~, \label{4D1}
\end{align}
which shows that Carter's constant is a quadrature with respect to the momentum of a particle.
A Killing tensor is connected with not only integral of geodesic motion
but separability of the Hamilton-Jacobi equation for geodesics.
The relationship to the separability was investigated 
by Benenti and Francaviglia \cite{Benenti:1975, Benenti:1979}.

In 1973, Floyd \cite{Floyd:1973} pointed out that 
the Killing tensor of the Kerr spacetime can be obtained in the form
\begin{align}
K_{ab} = f_{ac}f_b{}^c ~, \label{4D2}
\end{align}
where $\bm{f}$ is a Killing-Yano tensor \cite{Bochner:1948,Yano:1952} obeying
\begin{align}
f_{ab}=f_{[ab]} ~, ~~~ \nabla_{(c}f_{a)b}=0 ~. \label{4D3}
\end{align}
A Killing-Yano tensor is in many aspects more fundamental than a Killing tensor.
Especially, having a Killing-Yano tensor 
one can always construct the corresponding Killing tensor using Eq.\ (\ref{4D2}).
On the other hand, not every Killing tensor can be decomposed 
in terms of a Killing-Yano tensor \cite{Collinson:1976,Stephani:1978}.
Penrose \cite{Penrose:1973} proved that 
the existence of such a tensor occurs only for the Kerr spacetime.
Moreover, Hughston and Sommers \cite{Hughston:1973} demonstrated that 
the Killing-Yano tensor generates two commuting Killing vectors 
corresponding isometries the Kerr spacetime originally has as follows:
\begin{align}
\xi^a \equiv \frac{1}{3}\nabla_b(*f)^{ba} ~,~~~
\eta^a \equiv K^a{}_b\xi^b ~. \label{4D4}
\end{align}
In this way, for the Kerr spacetime
all the symmetries necessary for complete integrability of the Hamilton-Jacobi equation for geodesics
can be generated by a single Killing-Yano tensor.

Let us see the explicit form of hidden symmetries of the Kerr spacetime.
In 1987, Carter \cite{Carter:1987} pointed out that 
the every rank-2 Killing-Yano tensors are obtained from a 1-form potential $\bm{b}$,
\begin{align}
\bm{f}=*d\bm{b} ~. \label{4D5}
\end{align}
Obviously, the Hodge dual $\bm{h}=*\bm{f}$ follows
\begin{align}
d\bm{h} =0 ~.
\end{align}
This 2-form $\bm{h}$ is called a {\it closed conformal Killing-Yano (CKY) tensor}.
For the metric (\ref{Kerr metric}), 
the 1-form potential $\bm{b}$ is given as
\begin{align}
\bm{b} = -\frac{1}{2}(r^2+a^2\sin^2\theta)dt+\frac{1}{2}a\sin^2\theta(r^2+a^2)d\phi ~,
\end{align}
which produces both KY tensor $\bm{f}$ and the closed CKY tensor $\bm{h}$ in the form
\begin{align}
\bm{f} = a\cos\theta\,\bm{e}^0\wedge\bm{e}^1+r\,\bm{e}^2\wedge\bm{e}^3 ~,~~~
\bm{h} = r\,\bm{e}^0\wedge\bm{e}^1+a\cos\theta\,\bm{e}^2\wedge\bm{e}^3 ~,
\end{align}
where $\{\bm{e}^\mu\}$ is the orthonormal basis given by Eq.\ (\ref{4DOB}).
Using Eq.\ (\ref{4D2}), we find the rank-2 irreducible Killing tensor is written as
\begin{align}
\bm{K} = a^2\cos^2\theta(\bm{e}^0\bm{e}^0-\bm{e}^1\bm{e}^1)+r^2(\bm{e}^2\bm{e}^2+\bm{e}^3\bm{e}^3) ~. \label{4DKT}
\end{align}
Since two Killing vectors are
\begin{align}
& \bm{\partial}_t = \frac{\sqrt{\triangle}}{\sqrt{\Sigma}}\,\bm{e}_0
                    +\frac{a\sin\theta}{\sqrt{\Sigma}}\bm{e}_2 ~, \label{KerrKVt} \\
& \bm{\partial}_\phi = -a \sin^2\theta\frac{\sqrt{\triangle}}{\sqrt{\Sigma}}\,\bm{e}_0
                    -\frac{(r^2+a^2)\sin\theta}{\sqrt{\Sigma}}\bm{e}_2 ~, \label{KerrKVp}
\end{align}
under a coordinate transformation
\begin{align}
\tau = t-a\phi ~,~~~ \sigma=\frac{\phi}{a} ~, \label{4DCT}
\end{align}
we obtain new Killing vectors $\bm{\partial}_\tau=\bm{\partial}_t$ and 
$\bm{\partial}_\sigma=-a^2\bm{\partial}_t-a\,\bm{\partial}_\phi$, i.e,
\begin{align}
\bm{\partial}_\sigma = -a^2\cos^2\theta\frac{\sqrt{\triangle}}{\sqrt{\Sigma}}\,\bm{e}_0
                           +\frac{r^2a\sin\theta}{\sqrt{\Sigma}}\bm{e}_2 ~,
\end{align}
which enable us to identify them as $\bm{\xi}=\bm{\partial}_\tau$ and 
$\bm{\eta}=\bm{\partial}_\sigma$ in Eq.\ (\ref{4D4}).
In addition to Eq.\ (\ref{4DCT}), the coordinate transformation
\begin{align}
p=a\cos\theta ~,
\end{align}
transforms the form of the metric (\ref{Kerr metric}) 
into a very simple algebraic form,
\begin{align}
ds^2 =& \frac{r^2+p^2}{P}dp^2
        +\frac{P}{r^2+p^2}(d\tau-r^2d\sigma)^2 \nonumber\\
      & +\frac{r^2+p^2}{Q}dr^2-\frac{Q}{r^2+p^2}(d\tau+p^2d\sigma)^2 ~,
\end{align}
where
\begin{align}
Q=r^2-2Mr+a^2 ~,~~~~ P=-p^2+a^2 ~.
\end{align}
This form of the Kerr metric was first used by Carter \cite{Carter:1968a} 
and later by Plebanski \cite{Plebanski:1975}.
The ``off-shell'' metric 
with $Q$ and $P$ replaced by arbitrary functions $Q(r)$ and $P(p)$ itself is of type D.
Higher-dimensional spacetimes with Killing-Yano symmetry 
naturally generalize this form of the metric.

It is known that the Kerr spacetime is 
of special algebraic type D of Petrov's classification \cite{Petrov:1954}.
All the vacuum type D solutions were derived in 1969 by Kinnersley \cite{Kinnersley:1969}.
The important family of type D spacetimes, including the black-hole metric like the Kerr metric, 
the metric describing the accelerating sources as the C-metric, 
or the non-expanding Kundt's class type D solutions, 
can be represented by the general seven-parameter metric
discovered by Plebanski and Demianski \cite{Plebanski:1976}.

\subsection{Underlying Structures}

\subsubsection{Separability theory of Hamilton-Jacobi equations}
For a $D$-dimensional manifold $({\cal M},\bm{g})$, a local coordinate system $q^M$ is called
a {\it separable coordinate system} if a Hamilton-Jacobi equation in these coordinates
\begin{align}
H(q^M,p_M) = \kappa_0 ~,~~~ p_M=\frac{\partial S}{\partial q^M} ~, \label{HJ eq}
\end{align}
where $\kappa_0$ is a constant, is integrable by (additive) separation of variables, i.e.,
\begin{align}
S = S_1(q^1,c)+S_2(q^2,c)+\cdots +S_D(q^D,c) ~, \label{add sep}
\end{align}
where $S_M(q^M,c)$ depends only on the corresponding coordinate $q^M$ 
and includes $D$ constants $c=(c_1,\cdots,c_D)$.
Separability structures of Hamilton-Jacobi equations have been studied
\cite{Woodhouse:1975,Benenti:1975,Benenti:1979,Kalnins:1981} since 1904, 
when Levi-Civita demonstrated that 
Hamilton-Jacobi equations are (additively) separable in the coordinates $q^M$ if and only if
\begin{align}
& \partial^M\partial^NH\,\partial_MH\,\partial_NH
  +\partial_M\partial_NH\,\partial_MH\,\partial_NH \nonumber\\
& ~~~~~-\partial^M\partial_NH\,\partial_MH\,\partial^NH
  -\partial_M\partial^NH\,\partial^MH\,\partial_NH=0 ~,~~~ (M\neq N,~\text{no sum}) \label{LC cond}
\end{align}
where $\partial_M=\partial/\partial q^M$, $\partial^M=\partial/\partial p_M$.
According to Benenti and Francavigrila \cite{Benenti:1979}, 
for each separability structure a family of separable coordinates exists
such that each coordinate system in this family admits 
a maximal number $r$ of ignorable coordinates where $0\leq r\leq D$.
Each system in this family is called a {\it normal separable coordinate system}
and the corresponding separability structure is fully characterized by such a family.
Given normal coordinates $q^M=(x_\mu,\psi_j)$ $(M=1,\cdots,D)$, 
Greek indices $\mu,\nu,\cdots$ ranging from $1$ to $D-r$ correspond to non-ignorable coordinates $x_\mu$
and Latin indices $j,k,\cdots$ ranging from $1$ to $r$ correspond to ignorable ones $\psi_j$, i.e.,
\begin{align}
\partial_j\,g^{MN}=0 ~.
\end{align}
Without loss of generality it is possible to prove that we write the metric in the form
\begin{align}
\left(\frac{\partial}{\partial s}\right)^2 
= \sum_{\mu=1}^{D-r}g^{\mu\mu}\left(\frac{\partial}{\partial x_\mu}\right)^2
  +\sum_{j,k=1}^rg^{jk}\frac{\partial}{\partial\psi_j}\frac{\partial}{\partial\psi_k} ~. \label{met form 3}
\end{align}
When we focus especially on the geodesic Hamiltonian
\begin{align}
H=g^{MN}p_Mp_N ~, \label{geo H}
\end{align}
by applying this to (\ref{LC cond}) together with (\ref{met form 3}), 
we obtain the differential equations for $g^{\mu\mu}$ and $g^{jk}$.
The general solutions of these equations are given 
by St\"ackel matrix $\phi$ and $\zeta$-matrices $\zeta_{(\mu)}$, 
which give the following form of the metric:
\begin{align}
& g^{\mu\mu}=\bar{\phi}^{\mu}{}_{(m)} ~,~~~ 
  g^{\mu M}=0 ~~~(a\neq\mu) ~, \nonumber\\
& g^{jk} = \sum_{\mu=1}^m \zeta_{(\mu)}^{jk}\bar{\phi}^{\mu}{}_{(m)} ~, \label{met sta}
\end{align}
where $m=D-r$, $\bar{\phi}^{\mu}{}_{(m)}$ ($\mu=1,\cdots,m$) is 
the $m$-th row of the inverse of a St\"ackel matrix $\phi$,
i.e., $\bar{\phi}^{\mu}{}_{(\rho)}\phi^{(\rho)}{}_{\nu}=\delta_{\mu\nu}$,
and $\zeta_{(\mu)}^{jk}$ is the ($j,k$)-element of a $\zeta$-matrix $\zeta_{(\mu)}$.
St\"ackel matrix is an $m\times m$ matrix such that each element $\phi^{(\mu)}{}_{\nu}$ depends only on $x_\nu$,
while $\zeta$-matrix $\zeta_{(\mu)}$ is a $r\times r$ matrix such that all the elements are 
functions depending only on $x_\mu$.
Additionally, it is shown that $D-r$ rank-$2$ Killing tensors obeying (\ref{4D1}) exist
such that their contravariant components can be written in the form
\begin{align}
& K_{(\nu)}^{\mu\mu}=\bar{\phi}^{\mu}{}_{(\nu)} ~,~~~ 
  K_{(\nu)}^{\mu M}=0 ~~~(M\neq\mu) ~, \nonumber\\
& K_{(\nu)}^{jk} = \sum_{\mu=1}^m \zeta_{(\mu)}^{jk}\bar{\phi}^{\mu}{}_{(\nu)} ~, \label{kt sta}
\end{align}
where the metric is included as $m$-th Killing tensor, i.e., $\bm{K}^{(m)}=\bm{g}$.

In order to understand the mechanism of separability,
there is a geometrical characterization of separability structure
described by Benenti.
This characterization is stated as the following theorem\cite{Benenti:1979}:\\[-0.3cm]

\begin{theorem} \label{Benenti}
A $D$-dimensional manifold $({\cal M},\bm{g})$ admits a separability structure
if and only if the following conditions hold:\\[-0.3cm]
\begin{enumerate}
\item There exist $r$ independent commuting Killing vectors $\bm{X}_{(j)}$
\begin{align}
[\bm{X}_{(j)},\bm{X}_{(k)}]=0 ~,
\end{align}
\item There exist $D-r$ independent Killing tensors $K_{(\mu)}$, 
which satisfy the relations
\begin{align}
[\bm{K}_{(\mu)},\bm{K}_{(\nu)}]=0 ~,~~~ 
[\bm{X}_{(j)},\bm{K}_{(\mu)}]=0 ~,
\end{align}
\item The Killing tensors $\bm{K}_{(\mu)}$ have in common $D-r$ eigenvectors $\bm{X}_{(\mu)}$ such that
\begin{align}
[\bm{X}_{(\mu)},\bm{X}_{(\nu)}]=0 ~,~~~ 
[\bm{X}_{(\mu)},\bm{X}_{(j)}]=0 ~,~~~ 
g(\bm{X}_{(\mu)},\bm{X}_{(j)})=0 ~,
\end{align}
\end{enumerate}
where through this theorem, the precise meaning of independence and commutators is that
the $r$ linear first integrals associated with the Killing vectors $\bm{X}_{(j)}$
and the $D-r$ quadratic first integrals associated with the Killing tensors $\bm{K}_{(\mu)}$
are functionally independent and commute with respect to Poisson brackets, respectively.
These commutators are called Schouten-Nijenhuis brackets 
\cite{Schouten:1940}.
\end{theorem}\quad

We know that for the Kerr metric, 
the Hamilton-Jacobi equation for geodesics gives rise to separation of variables.
As is expected, the Kerr spacetime possesses the separability structure of $r=2$.
Actually, we can easily see that the Killing vectors (\ref{KerrKVt}), (\ref{KerrKVp}) 
and the Killing tensor (\ref{4DKT}) satisfy the conditions in theorem \ref{Benenti}. 
All the constants appearing when the Hamilton-Jacobi equation is separated 
are associated with these symmetries.
We would like to emphasize that in the Kerr spacetime,
a single rank-2 Killing-Yano tensor generates these symmetries 
and fully characterizes the separability structure of the Kerr spacetime
\cite{Collinson:1974,Dietz:1977,Demianski:1980,Dietz:1981,Dietz:1982}.

\subsubsection{Symmetry operators}
It is very powerful to consider symmetry operators, 
which help us to understand a geometrical meaning of separation constants for Klein-Gordon and Dirac equations.
Originally, a symmetry operator is a differential operator 
introduced as a symmetry of differential equations.
For a differential operator ${\cal O}_1$,
a differential operator ${\cal O}_2$ is called a {\it symmetry operator} for ${\cal O}_1$
if they commute, i.e., $[{\cal O}_1,{\cal O}_2]=0$.
The existence of such operators implies counterparts of separation constants
in a differential equation.

Regarding Klein-Gordon equations, 
Carter \cite{Carter:1977} pointed out first that
given an isometry $\bm{\xi}$ and/or Killing tensor $\bm{K}$ 
one can construct the operator
\begin{align}
\hat{\xi} \equiv \xi^a\nabla_a ~,~~~
\hat{K} \equiv \nabla_aK^{ab}\nabla_b ~,
\end{align}
which gives the commutator with the scalar laplacian $\Box\equiv \nabla_ag^{ab}\nabla_b$ as
\begin{align}
[\Box, \hat{\xi}]=0 ~,~~~
[\Box, \hat{K}]=\frac{4}{3}\nabla_a(K_c{}^{[a}R^{b]c})\nabla_b ~.
\end{align}
It was demonstrated later by Carter and McLenaghan \cite{Carter:1979} 
that the second equation automatically vanishes whenever 
the Killing tensor is a square of a Killing--Yano tensor of arbitrary rank. 
Since the Killing tensor in the Kerr spacetime is generated by the Killing-Yano tensor,
the symmetry operator constructed from the Killing tensor commutes with the laplacian.
Then, for the Kerr spacetime there exist three symmetry operators for the laplacian, 
in which two of them are associated with two isometries.
Moreover, it is shown that these operators commute between themselves in the Kerr spacetime.
The existence of such operators implies separation of variables in the Klein-Gordon equation.
Carter and McLenaghan further found that an operator
\begin{align}
\hat{f} \equiv i\gamma_5\gamma^a\Big(f_a{}^b\nabla_b-\frac{1}{6}\gamma^b\gamma^c\nabla_cf_{ab}\Big)
\end{align}
commutes the Dirac operator $\gamma^a\nabla_a$
whenever $\bm{f}$ is a Killing--Yano tensor.
Similarly, symmetry operators for other equations with spin,
including electromagnetic and gravitational perturbations were discussed.
\cite{Kamran:1984,Kamran:1985,Castillo:1988,Kalnins:1990
}

%%%%%%%%%%%%%%%%%%%%%%%%%%%%%%%%%%%%%%%%%%%%%%%%%%%
%%%%%%%%%%%%%%%%%%%%%%%%%%%%%%%%%%%%%%%%%%%%%%%%%%%
\section{Symmetry of Higher-Dimensional Spacetime}
%%%%%%%%%%%%%%%%%%%%%%%%%%%%%%%%%%%%%%%%%%%%%%%%%%%
Higher-dimensional solutions describing rotating black holes 
attract attention in the recent developments of supergravity and superstring theories.
Here, we focus especially on higher-dimensional rotating black hole spacetimes 
which are generalizations of the Kerr geometry.
In the four-dimensional Kerr spacetime, we saw that  
all the symmetries necessary for separability of the geodesic,
Klein-Gordon and Dirac equations, are described by the Killing-Yano (KY) tensor, 
or equivalently by the closed conformal Killing-Yano (CKY) tensor.
On the other hand, one would find that in higher dimensions, 
a  closed CKY tensor more crucially works than a KY tensor.
The purpose of this section is to introduce a notion of a CKY tensor on higher-dimensional spacetimes 
and to clarify its relationship to the integrability of geodesic equations. 

\subsection{Killing tensor and conformal Killing-Yano tensor}

\subsubsection{Killing and conformal Killing vector}
The geodesic motion of a particle is described by the geodesic equation
\begin{align}
p^b\nabla_bp^a = 0 ~, \label{GE}
\end{align}
where $\bm{p}$ represents a tangent to the geodesicD
Since it is difficult in general to solve this equation, 
we ordinarily study the constants of motion and simplify the discussion.
For a vector $\bm{k}$, we consider a inner product $k_ap^a$
which is the simplest invariant constructed from $\bm{k}$ and $\bm{p}$.
If this quantity is a constant along the geodesic, 
then the equation
\begin{align} \label{Eq2_1}
p^b\nabla_b(k_ap^a) = 0 
\end{align}
must be satisfied.
Since the left hand side of (\ref{Eq2_1}) is calculated as
$k_ap^b\nabla_bp^a + p^ap^b\nabla_{(b} k_{a)}$ 
and then the first term vanishes by the geodesic equation, 
a vector $\bm{k}$ must obey the equation
\begin{align}
\nabla_{(a}k_{b)}=0 \label{Killing equation} ~.
\end{align}
This vector field is called a Killing vector.

In above discussionC
we defined Killing vector from the view point of constants of motion along geodesics 
where Eq.\ (\ref{Eq2_1}) is essence of the discussion.
Now we shall consider a null geodesic.
Since we have $g_{ab}p^ap^b=0$,
we may add the term which is proportional to the metric
into the right hand side of (\ref{Eq2_1}).
Thus $\nabla_{(a}k_{b)}\propto g_{ab}$ is 
condition that the inner product $k_ap^a$ is constant along the null geodesic.
This is a definition of conformal Killing vector $\bm{k}$:~if there exists a function $q$ such that
\begin{align}
\nabla_{(a} k_{b)}=q g_{ab} ~,
\end{align}
then $\bm{k}$ is called a conformal Killing vector, and the inner product $k_ap^a$
becomes constant along geodesic.

There are two generalizations of Killing vectors to higher-rank tensors.
One is generalization to symmetric tensors and another is to anti-symmetric tensors.
These tensors are called Killing tensors and Killing-Yano tensors, respectively.
There are similar ways to generalize the conformal Killing vector 
and they are called conformal Killing tensors 
and conformal Killing-Yano tensors.

\begin{table}[t]
\begin{center}
\begin{tabular}{lll} \hline  & & \\[-0.3cm]
vectors &Killing vector &conformal Killing vector \\ & & \\[-0.3cm] \hline\hline & & \\[-0.3cm]
symmetric tensors &Killing tensor &conformal Killing tensor \\ & & \\[-0.3cm]
anti-symmetric tensors &Killing-Yano (KY) tensor  &conformal Killing-Yano (CKY) tensor \\ & & \\[-0.3cm] \hline
\end{tabular}
\caption{The generalizations of Killing and conformal Killing vectors}
\end{center}
\end{table}

\subsubsection{Killing tensor}
We consider  Killing tensors which are  symmetric generalizations of the Killing vector.
As previous arguments, we define Killing tensors from the view point of constants of motion along geodesic.
For a rank-$p$ symmetric tensor, i.e., $K_{(a_1\dots a_p)}=K_{a_1\dots a_p}$, 
the condition that a quantity $K_{a_1\dots a_p}p^{a_1}\cdots p^{a_p}$ 
is a constant along the geodesic requires 
\begin{align} \label{CKC}
p^b\nabla_b(K_{a_1\dots a_p}p^{a_1}\cdots p^{a_p})=0 ~,
\end{align}
where $\bm{p}$ is a tangent to the geodesic.
By using the geodesic equation, since the left hand side becomes 
$p^{a_1}\cdots p^{a_p}p^b\nabla_{(b}K_{a_1\dots a_p)}$, the equation
$\nabla_{(b}K_{a_1\dots a_p)} = 0$ is the condition that is necessary to satisfy
the equation (\ref{CKC}). 
A rank-$p$ Killing-tensor $\bm{K}$ is a symmetric tensor obeying the equation
\begin{align} \label{K_1}
\nabla_{(b}K_{a_1\dots a_p)} = 0 ~.
\end{align}
If $\bm{K}$ is a Killing tensor and $\bm{p}$ is a geodesic tangent,
then the quantity $K_{a_1\dots a_p}p^{a_1}\cdots p^{a_p}$ 
is a constant along the geodesic.

As conformal Killing vectors, in null geodesic case,
we can add the quantity which is proportional to $g_{ab}p^ap^b$ 
into the right-hand side of (\ref{CKC}).
Noting a symmetry of the indices, we have
\begin{align}
p^bp^{a_1}\cdots p^{a_p}\nabla_{(b}K_{a_1\dots a_p)} 
= p^bp^{a_1}\cdots p^{a_p}g_{(ba_1}Q_{a_2\dots a_p)} ~.
\end{align}
If there exists a rank-$(p-1)$ tensor $Q_{a_2\dots a_p}$ such that 
\begin{align} \label{CK_1}
\nabla_{(b}K_{a_1\dots a_p)} = g_{(ba_1}Q_{a_2\dots a_p)} ~,
\end{align}
then $\bm{K}$ is called a rank-$p$ conformal Killing tensor.

\subsubsection{Conformal Killing-Yano tensor}
A rank-$p$ conformal Killing-Yano (CKY) tensor\footnote{
Yano \cite{Yano:1952} discussed an anti-symmetric tensor $\bm{h}$ obeying
$\nabla_{(a}h_{b)c_1 \cdots c_{p-1}}=g_{ab} \xi_{c_1 \cdots c_{p-1}}$
as a candidate of a CKY tensor. 
Unfortunately, this tensor represents a Killing-Yano tensor
because it is proved that $\bm{\xi}$ vanishes identically.}, 
as a generalization of a conformal Killing vector, 
was introduced by Tachibana \cite{Tachibana:1969}  for the case of rank-2 
and Kashiwada \cite{Kashiwada:1968} for arbitrary rank. This is an anti-symmetric tensor
$\bm{h}$  obeying
\begin{align} \label{CKY}
\nabla_{(a}h_{b)c_1\dots c_{n-1}} = 
g_{ab}\xi_{c_1\dots c_{n-1}}+\sum_{i=1}^{n-1}(-1)^ig_{c_i(a}\xi_{b)c_1\dots \hat{c_i}\dots c_{n-1}}~. 
\end{align}
By tracing both sides one obtains the expression 
\begin{align}\label{ascky}
\xi_{c_1 \cdots c_{p-1}}=\frac{1}{D-p+1}\nabla^a h_{ac_1 \cdots c_{p-1}} ~.
\end{align}
If a CKY tensor satisfies the condition $\bm{\xi}=0$, 
it reduces to a Killing-Yano tensor.
Since the right-hand side of (\ref{CKY}) vanishes, 
the Killing-Yano tensor is defined as follows:
an anti-symmetric tensor $\bm{f}$ satisfying
\begin{align} \label{KY}
\nabla_{(b}f_{a_1)a_2\dots a_p} = 0 ~,
\end{align}
is called a {\it rank-$p$ Killing-Yano (KY) tensor}.\\[-0.3cm]

\begin{prop} \label{Prop1OfCKY}
Let $\bm{h}$ be a rank-$p$ CKY tensor. 
Then, a rank-$2$ symmetric tensor $\bm{K}$ defined by
\begin{align}
K_{ab} = h_{ac_1\dots c_{p-1}}h_b{}^{c_1\dots c_{p-1}}
\end{align}
is a conformal Killing tensor.
In particular, $\bm{K}$ is a rank-2 Killing tensor if $\bm{h}$ is a KY tensor.\\
\end{prop}

We shall identify an anti-symmetric tensor $\bm{h}$ with the $p$-form
\begin{align}
\bm{h}=\frac{1}{p!}\,h_{a_1a_2 \cdots a_{p}}dx^{a_1} \wedge \cdots \wedge dx^{a_p}.
\end{align}
Then, the rank-$p$ CKY tensor obeys the equation \cite{Semmelmann:2002}
\begin{align} \label{CKY_DF}
\nabla_X \bm{h}
= \frac{1}{p+1} \bm{X}\hook d\bm{h}
  -\frac{1}{D-p+1}\bm{X}^* \wedge \delta \bm{h} ~
\end{align}
for all vector fields $\bm{X}$. Here $\bm{X}^*=X_a dx^a$ denotes the 1-form dual to $\bm{X}=X^a \partial_a$ 
and $\delta$ the adjoint of the exterior derivative $d$. 
The symbol $\hook$ is an inner product. 
A CKY tensor $\bm{h}$ obeying $d\bm{h}=0$ is called a {\it closed CKY tensor}. 
Also, a CKY tensor $\bm{f}$ obeying $\delta \bm{f}=0$ is a Killing-Yano (KY) tensor.
One can show the following basic properties: \cite{Krtous:2007a}\\[-0.2cm]

\begin{prop} \label{Prop3OfCKY}
The Hodge dual $*\bm{h}$ of a CKY tensor $\bm{h}$ is also a CKY tensor. In particular,
the Hodge dual of a closed CKY tensor is a KY tensor and vice versa.
\end{prop}\quad\\[-0.6cm]

\begin{prop} \label{Prop4OfCKY}
Let $\bm{h}_1$ and $\bm{h}_2$ be two closed CKY tensors. 
Then their wedge product $\bm{h}_3 = \bm{h}_1\wedge \bm{h}_2$ is also a closed CKY tensor.
\end{prop}

\subsection{Geodesic integrability}

Frolov, Krtou\v{s}, Kubiz\v{n}\'ak, Page and Vasudevan \cite{Page:2007,Krtous:2007a,Krtous:2007b} 
have shown a simple procedure to construct a family of rank-2 Killing tensors.
They considered a special class of $D$-dimensional spacetimes endowed with a rank-$2$ closed CKY tensor $\bm{h}$.
By the condition $d\bm{h}=0$, the CKY tensor obeys the equations
\begin{align}
\nabla_X \bm{h}=-\frac{1}{D-1}\bm{X}^* \wedge \delta \bm{h}
\end{align}
In the tensor notation the equation above reads
\begin{align} \label{ckyD}
\nabla_a h_{bc}= \xi_c g_{ab}-\xi_b g_{ac} ~,~~\xi_a = \frac{1}{D-1} \nabla^b h_{ba}.
\end{align} 
They defined a $2j$-form $\bm{h}^{(j)}$ as
\begin{align}
\bm{h}^{(j)} 
= \underbrace{\bm{h} \wedge \bm{h} \wedge \cdots \wedge \bm{h}}_{j}
= \frac{1}{(2j)!} h^{(j)}_{a_1 \dotsm a_{2j}}
  dx^{a_1} \wedge \cdots \wedge dx^{a_{2j}} ~,
\end{align}
where the wedge products are taken $j-1$ times such as $\bm{h}^{(0)}=1$, 
$\bm{h}^{(1)}=h$, $\bm{h}^{(2)}=\bm{h}\wedge \bm{h}$, $\cdots$.
If we put the dimension $D=2n+\varepsilon$, 
where $\varepsilon=0$ for even dimensions and $\varepsilon=1$ for odd dimensions, 
$\bm{h}^{(j)}$ are non-trivial only for $j=0,\cdots,n-1+\varepsilon$,
i.e., $\bm{h}^{(j)}=0$ for $j>n-1+\varepsilon$.
The components are written as
\begin{align} \label{Eq3_1}
h^{(j)}_{a_1 \cdots a_{2j}}
= \frac{(2j)!}{2^j} h_{[a_1 a_2} h_{a_3 a_4} \dotsm h_{a_{2j-1} a_{2j}]} ~.
\end{align}
Since the wedge product of two closed CKY tensors is again a closed CKY tensor by proposition \ref{Prop4OfCKY},
$\bm{h}^{(j)}$ are closed CKY tensors for all $j$.
The proposition \ref{Prop3OfCKY} shows that 
the Hodge dual of the closed CKY tensors $\bm{h}^{(j)}$ gives rise to 
the Killing-Yano tensors $\bm{f}^{(j)}=* \bm{h}^{(j)}$. 
Explicitly, one has
\begin{align}
\bm{f}^{(j)} 
= * \bm{h}^{(j)} 
= \frac{1}{(D-2j)!}f^{(j)}_{a_1 \cdots a_{D-2j}}
  dx^{a_1} \wedge \cdots \wedge dx^{a_{D-2j}} ~,
\end{align}
where
\begin{align} \label{Eq3_2}
f^{(j)}_{a_1 \dotsm a_{D-2j}}
= \frac{1}{(2j)!} \epsilon^{b_1 \dotsm b_{2j}}{}_{a_1 \dotsm a_{D-2j}}
  h^{(j)}_{b_1 \dotsm b_{2j}} ~.
\end{align}
For odd dimensions, 
since $\bm{h}^{(n)}$ is a rank-$2n$ closed CKY tensor, $\bm{f}^{(n)}$ is a Killing vector.
Given these KY tensors $\bm{f}^{(j)}$ ($j=0,\dots,n-1$), 
one can construct the rank-2 Killing tensors 
\begin{align} \label{Eq3_3}
K^{(j)}_{ab}
= \frac{1}{(D-2j-1)!(j!)^2}
  f^{(j)}_{a c_1 \cdots c_{D-2j-1}} f_b^{(j)c_1 \cdots c_{D-2j-1}} ~, 
\end{align}
obeying the equation $\nabla_{(a}K^{(j)}_{bc)}=0$, and\\

\begin{prop} \label{Prop5OfCKY}
Killing tensors $\bm{K}^{(i)}$ are mutually commuting \cite{Page:2007,Krtous:2007a,Houri:2008a}:
\begin{align} \label{Eq3_4}
[\bm{K}^{(i)},\bm{K}^{(j)}] = 0~.
\end{align}
The bracket above represents the Schouten-Nijenhuis bracket. 
Hence, equivalently, the equation \eqref{Eq3_4} can be written as
\begin{align} \label{Eq3_5}
K^{(i)}_{d(a} \nabla^{d} K^{(j)}_{bc)}-K^{(j)}_{d(a} \nabla^{d} K^{(i)}_{bc)}=0 ~.
\end{align}
\end{prop}\quad

Furthermore, a family of Killing vectors was obtained from the rank-$2$ closed CKY tensor $\bm{h}$. 
\cite{Houri:2008a}
We first note the following two properties :
\begin{align} \label{assumption}
 (\textrm{a})~ \pounds_{\xi} \,\bm{g}=0 ~,~~
(\textrm{b})~ \pounds_{\xi} \,\bm{h}=0 ~,
\end{align}
where $\bm{\xi}$ is the associated vector of $\bm{h}$ defined by \eqref{ckyD}.
Actually, as we will see in section 5.1, 
the conditions (\,a\,) and (\,b\,) follow from the existence of the closed CKY tensor.
From (\,a\,) we have 
$\pounds_{\xi} \ast\bm{h}^{(j)}=\ast\, \pounds_{\xi}\,\bm{h}^{(j)}$ 
and hence
(\,b\,) yields
\begin{align} \label{LieDer}
\pounds_{\xi} \bm{h}^{(j)}=0,~~\pounds_{\xi} \bm{f}^{(j)}=0,~~ \pounds_{\xi} \bm{K}^{(j)}=0.
\end{align}
We also immediately obtain from \eqref{ckyD}
\begin{align} \label{CovD}
\nabla_{\xi} \bm{h}^{(j)}=0,~~\nabla_{\xi} \bm{f}^{(j)}=0,~~ \nabla_{\xi} \bm{K}^{(j)}=0.
\end{align}
Let us define the vector fields $\bm{\eta}^{(j)}~(j=0, \cdots , n-1) $~,
\begin{align}
\eta^{(j)}_a = K^{(j)}{}_{a}{}^b\xi_b. 
\end{align}
and $\bm{\eta}^{(n)}\equiv * \bm{f}^{(n)}$ for $\varepsilon=1$.
Then we have
\begin{align}
\nabla_{(a}\eta^{(j)}_{b)} =
\frac{1}{2}\pounds_{\xi}K^{(j)}_{ab}-\nabla_{\xi}K^{(j)}_{ab},
\end{align}
which vanishes by \eqref{LieDer} and \eqref{CovD},~i.e. 
$\bm{\eta}^{(j)}~(j=0, \cdots, n-1+\varepsilon)$ are Killing vectors\footnote{
These Killing vectors $\bm{\eta}^{(j)}$ can be also written as 
$\bm{\eta}^{(j)}=-\delta\bm{\omega}^{(j)}$ by the Killing co-potential\cite{CveticEtal:2010}
\begin{align}
\omega^{(j)}_{ab}=\frac{1}{n-2j-1}K^{(j)}_{c[a}h^c{}_{b]} ~~(j=0,\cdots,n-1) ~.
\end{align}
}.\\

\begin{prop} \label{Prop6OfCKY}
Killing vectors $\bm{\eta}^{(i)}$ and Killing tensors $\bm{K}^{(j)}$
are mutually commuting\cite{Houri:2008a},
\begin{align}
[\bm{\eta}^{(i)},\bm{K}^{(j)}] =0 ~,~~ [\bm{\eta}^{(i)},\bm{\eta}^{(j)}] =0 ~.
\end{align}
\end{prop}

In general, $\bm{K}^{(i)} ~(i=0,\cdots, n-1)$ and 
$\bm{\eta}^{(j)}~(j=0,\cdots, n-1+\varepsilon)$ are not independent. 
In section 5.2, we will see that on a $(2n+\varepsilon)$-dimensional spacetime 
with a rank-2 closed CKY tensor $\bm{h}$
there is a pair of integers $(\ell, \ell+\delta)$ with $0 \le \ell \le n$ and $\delta=0,~1$ 
such that both $\bm{K}^{(i)} ~(i=0,\cdots, \ell-1)$ 
and $\bm{\eta}^{(j)}~(j=0,\cdots, \ell-1+\delta )$ are independent.
We call $(\ell,\ell+\delta)$ the order of $h$, 
and a rank-2 closed CKY tensor with the maximal order $(n, n+\varepsilon)$ 
a principal conformal Killing-Yano (PCKY) tensor \cite{Krtous:2007a,Frolov:2008c}.
Propositions \ref{Prop5OfCKY} and \ref{Prop6OfCKY} mean that 
the geodesic equations on spacetimes admitting a PCKY tensor are completely integrable in the Liouville sense.

%%%%%%%%%%%%%%%%%%%%%%%%%%%%%%%%%%%%%%%%%%%%%%%%%%
%%%%%%%%%%%%%%%%%%%%%%%%%%%%%%%%%%%%%%%%%%%%%%%%%%
%%%%%%%%%%%%%%%%%%%%%%%%%%%%%%%%%%%%%%%%%%%%%%%%%%
\section{Higher-Dimensional Kerr Geometry}
The Kerr-NUT-(A)dS spacetimes describe the most general rotating black holes with spherical horizon. 
We review these spacetimes from the view point of symmetry, and
also present the explicit separation of variables in Hamilton-Jacobi, Klein-Gordon and Dirac equations.

\subsection{Kerr-NUT-(A)dS spacetimes}
We start with a class of $D$-dimensional metrics found by Chen, L\"{u} and Pope\cite{Chen:2006b}.
The metrics are written in the local coordinates $q^M=(x_\mu,\psi_k)$ 
where the latter coordinates $\psi_k$ represent the isometries of spacetime. 
They are explicitly given as follows:\\
(a) $D=2 n$ 
\begin{align}\label{KerrNUT1}
\bm{g} = \sum_{\mu=1}^{n} \frac{U_\mu}{X_\mu} dx_{\mu}^2+
\sum_{\mu=1}^{n} \frac{X_{\mu}}{U_\mu} \left( \sum_{k=0}^{n-1} A_{\mu}^{(k)}
 d\psi_k \right)^2~~, 
\end{align}
(b) $D=2 n+1$
\begin{align}\label{KerrNUT2}
\bm{g} = \sum_{\mu=1}^{n}\frac{U_\mu}{X_\mu} dx_{\mu}^2+
\sum_{\mu=1}^{n} \frac{X_{\mu}}{U_\mu} \left( \sum_{k=0}^{n-1} A_{\mu}^{(k)}
 d\psi_k \right)^2+\frac{c}{A^{(n)}}\left( \sum_{k=0}^{n} A^{(k)}
 d\psi_k \right)^2~~. 
\end{align}
The metric functions are defined by
\begin{align}\label{yu}
& U_{\mu} = \prod_{\substack{\nu=1\\(\nu \neq \mu)}}^n
            ( x_{\mu}^2 - x_{\nu}^2 ) ~,~~~
  A^{(k)} = \sum_{1\leq\nu_1<\nu_2<\cdots<\nu_k\leq n}
            x_{\nu_1}^2 x_{\nu_2}^2 \cdots x_{\nu_k}^2 ~, \nonumber\\
& A^{(k)}_\mu = \sum_{\substack{1 \leq \nu_1 < \nu_2 < \dotsm
< \nu_k \leq n \\(\nu_i \neq \mu)}}
x_{\nu_1}^2 x_{\nu_2}^2 \dotsm x_{\nu_k}^2,
\end{align}
with a constant $c$ and $X_\mu$ is an arbitrary function depending only on $x_\mu$.
It is worth remarking that $A^{(k)}$ and $A^{(k)}_\mu$ 
are the elementary symmetric functions of $x_\nu^2$'s defined via the generating functions
\begin{eqnarray}
\prod_{\nu=1}^n(t-x_\nu^2)=t^n-A^{(1)} t^{n-1}+ \cdots + (-1)^n A^{(n)},\nonumber\\
\prod_{\nu \ne \mu}(t-x_\nu^2)=t^{n-1}-A^{(1)}_\mu t^{n-2}+ \cdots + (-1)^{n-1} A^{(n-1)}_\mu.
\end{eqnarray}
To treat both cases of even and odd dimensions simultaneously we denote
\begin{align}
D=2n+\varepsilon,
\end{align}
where $\varepsilon=0$ and $\varepsilon=1$ for even and odd number of dimensions, respectively.
We shall use the following orthonormal basis for the metric
\begin{align}
\bm{g}=\sum_{\mu=1}^n(\bm{e}^\mu\bm{e}^\mu+\bm{e}^{\hat{\mu}}\bm{e}^{\hat{\mu}})+\varepsilon\,\bm{e}^0\bm{e}^0 ~,
\end{align}
where
\begin{align}\label{ortho1}
\bm{e}^{\mu} = \sqrt{\frac{U_{\mu}}{X_\mu}}dx_\mu, \quad
\bm{e}^{\hat{\mu}} = \sqrt{\frac{X_{\mu}}{U_\mu}}
\left( \sum_{k=0}^{n-1} A_{\mu}^{(k)}  d\psi_k \right).
\end{align}
In odd-dimensional case we add a 1-form
\begin{align}\label{ortho2}
\bm{e}^0 = \sqrt{\frac{c}{A^{(n)}}}
\left( \sum_{k=0}^{n} A^{(k)} d\psi_k \right).
\end{align}

The metric admits a rank-2 closed CKY tensor\cite{Frolov:2007,Kubiznak:2007}
\begin{align}\label{pcky}
\bm{h}=\frac{1}{2}\sum_{k=0}^{n-1}dA^{(k+1)} \wedge d\psi_k
= \sum_{\mu=1}^n x_\mu \,\bm{e}^\mu \wedge \bm{e}^{\hat{\mu}}.
\end{align}
According to section 3.2 the associated Killing tensors $\bm{K}^{(j)} (j=0, \cdots, n-1)$ and Killing vectors $\bm{\eta}^{(j)} (j=0, \cdots, n-1+\varepsilon)$ are calculated as
\begin{align}
\bm{K}^{(j)}=
  \sum_{\mu=1}^nA_\mu^{(j)}(\bm{e}^\mu\bm{e}^\mu+\bm{e}^{\hat{\mu}}\bm{e}^{\hat{\mu}})
  +\varepsilon\,A^{(j)}\,\bm{e}^0\bm{e}^0 ~,~~~
\bm{\eta}^{(j)}=\frac{\partial}{\partial \psi_j} ~. \label{HDKT}
\end{align}
These quantities are clearly independent and 
hence the CKY tensor $\bm{h}$ has the maximal order $(n,n+\varepsilon)$, i.e., 
$\bm{h}$ is a PCKY tensor. 
Thus, the geodesic equations are completely integrable.

The spacetime has locally two orthogonal foliations $\{W_{n+\varepsilon} \}$ and $\{Z_n \}$ \cite{Benenti:1979}, 
where each integral submanifold $W_{n+\varepsilon}$ is flat in the induced metric 
and each $Z_{n}$ is a totally geodesic submanifold. 
The foliation $\{W_{n+\varepsilon} \}$ is actually that of the integral submanifolds 
associated with the involutive distribution $\{ \bm{\eta}^{(j)}=\partial/\partial \psi_j \}$,
while the foliation $\{Z_n \}$ is the complementary foliation 
associated with the involutive distribution $\{ \bm{v}_\mu=\partial/\partial x_\mu \}$, 
where $\bm{v}_\mu$ are in common $n$ eigenvectors of the Killing tensors,
$\bm{K}^{(j)}\cdot\bm{v}_\mu=A^{(j)}_\mu \bm{v}_\mu$.
The foliations associated with complex structure were also discussed by Mason and Taghavi-Chabert.
\cite{Mason:2010}

The metrics \eqref{KerrNUT1} and \eqref{KerrNUT2} satisfy the Einstein equations 
\begin{align}
R_{ab}= \Lambda g_{ab}
\end{align}
if and only if the metric functions $X_\mu$ take the form \cite{Hamamoto:2007}:
\begin{align} \label{Einstein1}
X_{\mu}=\sum_{k=0}^{n} c_{2 k} x_{\mu}^{2 k}+b_{\mu} x_{\mu}^{1-\varepsilon}
+\varepsilon \frac{(-1)^{n}c}{x_{\mu}^2} , 
\end{align}
where $c, c_{2 k}$ and $b_{\mu}$ are
free parameters and $\Lambda=-(D-1)c_{2n}$. 
The Einstein metric given by \eqref{Einstein1} describes the most general known
higher-dimensional rotating black hole solution with NUT parameters
in an asymptotically (A)dS spacetime, i.e.,
Kerr-NUT-(A)dS metric\footnote{
In this paper we concentrate only on 
the class of rotating black holes with spherical horizon topology. 
It is known that in higher-dimensions
there exist different type of rotating black objects such as black rings 
and their generalizations. \cite{Emparan:2008}
These black objects do not belong to the class of the Kerr-NUT-(A)dS metrics. 
}. 
Then, the parameters  $c, c_{2 k}$ and $b_{\mu}$ are related to rotation parameters, mass, and NUT parameters.   
The higher-dimensional vacuum rotating black hole solutions
discovered by Myers and Perry \cite{Myers:1986} and 
by Gibbons, L\"u, Page, and Pope \cite{Gibbons:2004a,Gibbons:2005}
are recovered when the some parameters vanish (see Table II).
The existence of the PCKY tensor is irrelevant 
whether the metrics satisfy the Einstein equations, 
so that it is possible to consider the separation of variables 
in a broad class of metrics where $X_\mu$'s are arbitrary functions of one variable $x_\mu$.
In addition, it was shown that this class is of the algebraic type D 
of the higher-dimensional classification.
\cite{Hamamoto:2007,Coley:2004,Milson:2005,Pravda:2007,Coley:2008,Mason:2010}

The Kerr-NUT-(A)dS spacetimes possess a PCKY tensor \eqref{pcky},
which generalizes four-dimensional Kerr geometry to higher dimensions; 
the separation of Hamilton-Jacobi, Klein-Gordon and Dirac equations. 
It is interesting to ask whether separations of other field euations are generalized.
While it is known that in four dimensions there is a link between separation of Maxwell equations
and the existence of a Killing-Yano tensor, 
such a link has never been shown and even separation of the equations is not known.

\begin{table}[t]
\begin{center}
\begin{tabular}{|c|c|c|c|c|c|}
\hline @
\quad   & mass &  rotation & NUT & $\Lambda$ & parameter \\
\hline\hline
Myers-Perry~(1986) & $\bigcirc$ & $\bigcirc$& $ \times$& 0 & 1+[(D-1)/2] \\
\hline
Gibbons-L\"u-Page-Pope~(2004) & $\bigcirc$& $\bigcirc$& $\times$ &non-zero & 2+[(D-1)/2] \\
\hline
Chen-L\"u-Pope~(2006) & $\bigcirc$& $\bigcirc$& $\bigcirc$ & non-zero & D \\
\hline
\end{tabular}
\caption{$D$-dimensional rotating black hole solutions with spherical horizon topology}
\end{center}
\end{table}

\subsection{Separation of variables}
\subsubsection{Separability of the Hamilton-Jacobi equation}
It was shown \cite{Frolov:2003,Frolov:2007b,Chong:2005,Kunduri:2005,Davis:2006} 
that in the Kerr-NUT-(A)dS spacetime the Hamilton-Jacobi equation for geodesics,
\begin{align}\label{HJ}
\frac{\partial S}{\partial \lambda}+g^{MN}\partial_M S\,\partial_N S=0 ~,
\end{align}
allows an additive separation of variables 
\begin{align}
S=-\kappa_0\lambda+\sum_{\mu=1}^n S_{\mu}(x_{\mu})+ \sum_{k=0}^{n-1+\varepsilon} n_k\psi_k ~, \label{sepS}
\end{align}
with functions ${S_\mu(x_\mu)}$ of a single argument ${x_\mu}$ and 
constants $\kappa_0$ and $n_k$.
Following section 2.2.1, we shall review this separation briefly.
For the separated solution (\ref{sepS}), 
the equation (\ref{HJ}) reduces to the geodesic Hamilton--Jacobi equation
\begin{align}
g^{MN}p_Mp_N = \kappa_0 ~,~~~ 
p_\mu=\frac{d S_\mu}{d x_\mu} ~,~~~
p_k = n_k ~,
\end{align}
which can be regarded as one of the differential equations
\begin{align}\label{HJ2}
K^{MN}_{(j)}p_Mp_N = \kappa_j ~.
\end{align}
From (\ref{HDKT}), we find that in the standard form \eqref{kt sta} $\bm{K}^{(j)}$ are written as
\begin{align}
& K^{\mu \mu}_{(j)}=\bar{\phi}^{\mu}{}_{(j)} ~,~~~
  K^{\mu M}_{(j)}=0 ~,~~~ (M\neq\mu) \nonumber\\
& K^{k \ell}_{(j)}=\sum_{\mu=1}^n \zeta^{k \ell}_{(\mu)}\bar{\phi}^{\mu}{}_{(j)} ~,
\end{align}
where the St\"ackel matrix and $\zeta$-matrices are given by
\begin{align}
& \phi^{(j)}{}_{\mu}= \frac{(-1)^jx_\mu^{2(n-j-1)}}{X_\mu} ~,~~~
  \bar{\phi}^{\mu}{}_{(j)}=\frac{A^{(j)}_\mu X_\mu}{U_\mu} ~, \label{HDstackel1} \\
& \zeta^{k \ell}_{(\mu)}
  =\frac{(-1)^{k+\ell} x_\mu^{2(2n-k-\ell-2)}}{X_\mu^2}
  +\frac{(-1)^{n+1}}{c x_\mu^2 X_\mu} \delta_{n k} \delta_{n \ell} ~. \label{HDstacekl2}
\end{align}
Now, the equation \eqref{HJ2} takes the form
\begin{align}
\sum_{\mu=1}^n \bar{\phi}^\mu_{(j)} p_\mu^2+\sum_{\mu=1}^n \sum_{k,\ell=0}^{n-1+\varepsilon} \zeta^{k \ell}_{(\mu)}\bar{\phi}^\mu_{(j)}n_k n_\ell=\kappa_j,
\end{align}
which can be solved with respect to the momenta $p_\mu=dS_\mu/dx_\mu$,
\begin{align}
\left(\frac{dS_\mu}{dx_\mu} \right)^2=\sum_{j=1}^n\phi^{(j)}{}_{\mu}\kappa_j
-\sum_{k,\ell=1}^{n-1-\varepsilon}\zeta^{k\ell}_{(\mu)}n_k n_\ell~.
\end{align}
Since both $\phi^{(j)}{}_{\mu}$ and $\zeta^{k\ell}_{(\mu)}$ depend on $x_\mu$ only, the functions $S_\mu$ are given by
\begin{align} \label{solutionHJ}
S_\mu(x_\mu)=\int \Bigl(\sum_{j=0}^{n-1+\varepsilon}\phi^{(j)}{}_{\mu} \kappa_j
-\sum_{k,\ell=0}^{n-1+\varepsilon}\zeta^{k \ell}_{(\mu)}n_k n_\ell \Bigl)^{1/2}dx_\mu ~.
\end{align}

%%%%%%%%%%%%%%%%%%%%%%%%%%%%%%%%%%%%%%%%%%%%%%%%%%%%%%
\subsubsection{Separability of the Klein-Gordon equation}
The behavior of a massive scalar field $\Phi$ is governed by the Klein--Gordon equation
\begin{align}\label{KG}
\Box\Phi=\frac{1}{\sqrt{|g|}}\,\partial_{M}(\sqrt{|g|}g^{MN}\partial_{N}\Phi)=\mu^2\Phi.
\end{align}
According to the paper \cite{Frolov:2007b} 
one can demonstrate that this equation in the Kerr-NUT-(A)dS
background allows a multiplicative separation of variables
\begin{align}\label{multsep}
\Phi=\prod_{\mu=1}^nR_{\mu}(x_{\mu})\prod_{k=0}^{n-1+\varepsilon}e^{in_k \psi_k}.
\end{align}
Using the expression
\begin{align}
g^{\mu \mu}=\frac{X_\mu}{U_\mu}=\bar{\phi}^{\mu}{}_{(0)} ~,~~
g^{k \ell}=\sum_{\mu=1}^n \zeta^{k \ell}_{(\mu)}\bar{\phi}^{\mu}{}_{(0)}
\end{align} 
we have the following explicit form  
\begin{align}\label{KGm}
\sum_{\mu=1}^n \frac{1}{U_\mu}\Bigl(\partial_{\mu}X_\mu \partial_\mu \Phi+\frac{\varepsilon X_\mu \partial_\mu \Phi}{x_\mu}+
\sum_{k,\ell=0}^{n-1+\varepsilon} \zeta^{k \ell}_{\mu} X_\mu \partial_k \partial_\ell \Phi-\mu^2 x_\mu^{2(n-1)}\Phi \Bigl)=0.
\end{align}
We further notice that 
\begin{align}
\partial_{k}\, \Phi=in_k \, \Phi\,,\quad \partial_{\mu} \,\Phi=\frac{R_{\mu}'}{R_{\mu}} \, \Phi\,,\quad  \partial^2_{\mu} \,\Phi=\frac{R_{\mu}''}{R_{\mu}}~\Phi\,,
\end{align}
and then the Klein--Gordon equation takes the form
\begin{align}\label{KGsep}
\sum_{\mu=1}^n\frac{G_{\mu}}{U_\mu}\,\Phi=0\,,
\end{align}
where $G_{\mu}$ is a function of $x_{\mu}$ only,
\begin{align}
G_{\mu}=X_{\mu}\frac{R_{\mu}''}{R_{\mu}}+\frac{R'_{\mu}}{R_{\mu}}\Bigl(X_{\mu}'+\epsilon\frac{X_{\mu}}{x_{\mu}}\Bigr)
-\sum_{k,\ell=0}^{n-1+\varepsilon} \zeta^{k \ell}_\mu X_\mu n_k n_\ell-\mu^2 x_{\mu}^{2(n-1)}.
\end{align}
Here, the prime means the derivative of functions 
${R_\mu}$ and ${X_\mu}$ with respect to their single argument ${x_\mu}$.
If we use the identity
\begin{align}
\sum_{\mu=1}^n \frac{x_\mu^{2k}}{U_\mu}=0~~(k=0,1, \cdots, n-2),
\end{align}
then the general solution of (\ref{KGsep}) is given by
\begin{align}
G_{\mu}=-\sum_{j=1}^{n-1} (-1)^j\kappa_{j} x_\mu^{2(n-1-j)}\;
\end{align}
with arbitrary constants $\kappa_j$.

Therefore, we have demonstrated that the Klein--Gordon equation (\ref{KG}) in the 
background allows a multiplicative separation
of variables (\ref{multsep}), where functions $R_{\mu}(x_{\mu})$ satisfy the ordinary 
second order differential equations
\begin{align}\label{ODE}
R_{\mu}''+\Bigl(\frac{X_{\mu}'}{X_\mu}+\frac{\varepsilon}{x_{\mu}}\Bigr) R_{\mu}'+
\Bigl(\sum_{j=0}^{n-1}\phi_{\mu}^{(j)} \kappa_j-\sum_{k,\ell=0}^{n-1+\varepsilon}
\zeta^{k \ell}_{\mu}n_k n_\ell \Bigr)R_\mu =0 ,
\end{align}
where we have used the St\"ackel matrix $\phi_{\mu}^{(j)}$ \eqref{HDstackel1} together with $\kappa_0 \equiv -\mu^2$. It should be noted that
the last term exists in the solution \eqref{solutionHJ} of the Hamilton-Jacobi equation. This structure
can be naturally explained by considering the semiclassical solution of the Klein-Gordon equation. 
\cite{Sergyeyev:2008}

\subsubsection{Separability of the Dirac equation}
Finally, we demonstrate separability of the Dirac equation.\cite{Oota:2008}
The dual vector fields to Eq.\ \eqref{ortho1} and/or Eq.\ \eqref{ortho2} are given by
\begin{align}
\bm{e}_\mu=\sqrt{\frac{X_\mu}{U_\mu}} \frac{\partial}{\partial x_\mu},~~
\bm{e}_{\hat{\mu}}=\sum_{k=0}^{n-1+\varepsilon} \frac{(-1)^k x_\mu^{2(n-1-k)}}{\sqrt{X_\mu U_\mu}}\frac{\partial}{\partial \psi_k},~~
\bm{e}_{0}=\frac{1}{\sqrt{c A^{(n)}}}\frac{\partial}{\partial \psi_n}.
\end{align}
The corresponding connection 1-form $\bm{\omega}_{ab}$ is calculated as \eqref{connection}. Then, the Dirac equation is written in the form
\begin{align}\label{DiracCH}
\bigl(\gamma^MD_M+m\bigr)\Psi=0\,,\quad\  D_M=e_M+\frac{1}{4}\gamma^N\gamma^P\omega_{NP}(e_M)\,.
\end{align}

Let us use the following representation of $\gamma$-matrices: $\{ \gamma^a, \gamma^b \} = 2 \delta^{ab}$,
\begin{align} \label{EQ14}
\gamma^{\mu}=& 
\underbrace{\sigma_3 \otimes \sigma_3 \otimes \cdots \otimes \sigma_3}_{\mu-1}
\otimes \sigma_1 \otimes I \otimes \cdots \otimes I, \nonumber\\
\gamma^{\hat{\mu}}=& 
\underbrace{\sigma_3 \otimes \sigma_3 \otimes \cdots \otimes \sigma_3}_{\mu-1} 
\otimes \sigma_2 \otimes I \otimes \cdots \otimes I, \nonumber\\
\gamma^0=&\sigma_3 \otimes \sigma_3 \otimes \cdots \otimes \sigma_3\,,
\end{align} 
where $I$ is the $2 \times 2$ identity matrix and $\sigma_i$ are the Pauli matrices.
In this representation, we write the $2^n$ components of the spinor field as
$\Psi_{\epsilon_1 \epsilon_2  \cdots  \,\epsilon_n }~(\epsilon_{\mu}=\pm 1)$, and
it follows that
\begin{align} \label{EQ15}
(\gamma^{\mu}\Psi)_{\epsilon_1 \epsilon_2  \cdots  \epsilon_n }
=& \left(
\prod_{\nu=1}^{\mu-1}\epsilon_{\nu}\right)
\Psi_{\epsilon_1 \cdots  \epsilon_{\mu-1} (-\epsilon_{\mu})
\epsilon_{\mu+1} \cdots \epsilon_n },\nonumber\\
(\gamma^{\hat{\mu}}\Psi)_{\epsilon_1 \epsilon_2  \cdots  \epsilon_n } 
=& -i \epsilon_{\mu}
\left(\prod_{\nu=1}^{\mu-1} \epsilon_{\nu}\right)
\Psi_{\epsilon_1 \cdots  \epsilon_{\mu-1} (-\epsilon_{\mu})
\epsilon_{\mu+1} \cdots \epsilon_n },\nonumber\\
(\gamma^0\Psi)_{\epsilon_1 \epsilon_2  \cdots  \epsilon_n } 
=& \Big(\prod_{\rho=1}^n\epsilon_\rho\Big)
\Psi_{\epsilon_1 \cdots \epsilon_n }~.
\end{align}

We consider the separable solution
\begin{align}
\Psi = \hat{\Psi}(x)\exp \Big(i\sum_{k=0}^{n-1+\varepsilon}n_k\psi_k\Big) ~,
\end{align}
Using (\ref{EQ15}), we obtain the following explicit form
\begin{align}
& 
   \sum_{\mu=1}^n \sqrt{\frac{X_\mu}{U_\mu}}\Big(\prod_{\rho=1}^{\mu-1}\epsilon_\rho\Big)
    \Bigg[\frac{\partial}{\partial x_\mu}+\frac{X_\mu^\prime}{4X_\mu}
          +\frac{\varepsilon}{2x_\mu}+\frac{\epsilon_\mu Y_\mu}{X_\mu}          
          +\frac{1}{2}\sum_{\nu\neq\mu}\frac{1}{x_\mu-\epsilon_\mu\epsilon_\nu x_\nu}\Bigg] \,\hat{\Psi}_{\epsilon_1\dots\epsilon_{\mu-1}(-\epsilon_\mu)\epsilon_{\mu+1}\dots \epsilon_n} \nonumber\\
& ~~~+\Bigg[\varepsilon\,i\sqrt{\frac{c}{A^{(n)}}}\Big(\prod_{\rho=1}^{\mu-1}\epsilon_\rho\Big)
    \left(\frac{n_n}{c}
    -\sum_{\mu=1}^n\frac{\epsilon_\mu }{2 x_\mu}\right)\,
    +m \Bigg]\,\hat{\Psi}_{\epsilon_1\dots \epsilon_n} ~=~ 0 ~, \label{de1}
\end{align}
together with
\begin{align}
Y_\mu = \sum_{k=0}^{n-1+\varepsilon}(-1)^kx_\mu^{2(n-k-1)}n_k~.
\end{align}
Setting
\begin{align}
\hat{\Psi}_{\epsilon_1\dots \epsilon_n}(x) = 
\Bigg(\prod_{1\leq\mu<\nu\leq n}\frac{1}{\sqrt{x_\mu+\epsilon_\mu\epsilon_\nu x_\nu}}\Bigg)
\Bigg(\prod_{\mu=1}^n\chi^{(\mu)}_{\epsilon_\mu}(x_\mu)\Bigg) ~,
\end{align}
we have the equations
following from (\ref{de1}):
\begin{eqnarray}
& &\sum_{\mu=1}^n 
\frac{P^{(\mu)}_{\epsilon_\mu}(x_\mu)}{\prod_{\substack{\nu=1\\ \nu\neq\mu}}^n(\epsilon_\mu x_\mu-\epsilon_\nu x_\nu)}
+\frac{\varepsilon\,i\sqrt{c}}{\prod_{\rho=1}^n(\epsilon_\rho x_\rho)}
\Big(\frac{n_n}{c}-\sum_{\mu=1}^n\frac{\epsilon_\mu}{2x_\mu}\Big)+m=0 ~,\nonumber\\ \label{EQ20}
& &P^{(\mu)}_{\epsilon_\mu} = 
(-1)^{\mu-1}(\epsilon_\mu)^{n-\mu}\sqrt{(-1)^{\mu-1}X_\mu}\frac{1}{\chi^{(\mu)}_{\epsilon_\mu}}
\Big(\frac{d}{dx_\mu}+\frac{X_\mu^\prime}{4X_\mu}
+\frac{\epsilon_\mu Y_\mu}{X_\mu}\Big)\chi^{(\mu)}_{-\epsilon_\mu} ~. 
\end{eqnarray}
The functions $P^{(\mu)}_{\epsilon_\mu}$ depend on the variable $x_\mu$ only. 
In order to satisfy (\ref{EQ20}) $P^{(\mu)}_{\epsilon_\mu}$ must assume the form
$P^{(\mu)}_{\epsilon_\mu}(x_\mu) = Q(\epsilon_\mu x_\mu)\,$:
\begin{itemize}
\item[(a)] in an even dimension ($\varepsilon=0$)
\begin{align}
Q(y)=\sum_{j=0}^{n-1} q_j y^j\,,
\end{align}
\item[(b)] in an odd dimension ($\varepsilon=1$)
\begin{align}
Q(y)=\frac{q_{-2}}{y^2}+\frac{q_{-1}}{y}+\sum_{j=-2}^{n-1} q_j y^j ~,
\end{align}
where
\begin{align}
\ q_{n-1}=-m\,,\ \ q_{-1}=\frac{i}{\sqrt{c}}(-1)^n n_n\,,\ \ q_{-2}=\frac{i}{2} (-1)^{n-1} \sqrt{c}\,.
\end{align}
\end{itemize}
In both cases parameters $q_j$ ($j=0,\dots,n-2$) are arbitrary.
After all, we have proved that the
Dirac equation  in the Kerr-NUT-(A)dS spacetime  allows separation of variables 
\begin{align}
\Psi_{\epsilon_1\dots \epsilon_n} = 
\prod_{1\leq\mu<\nu\leq n}\frac{1}{\sqrt{x_\mu+\epsilon_\mu\epsilon_\nu x_\nu}}
\Bigg(\prod_{\mu=1}^n\chi^{(\mu)}_{\epsilon_\mu}(x_\mu)\Bigg)
\exp \Bigg(i\sum_{k=0}^{n-1+\varepsilon}n_k\psi_k\Bigg) ~,
\end{align}
where functions $\chi^{(\mu)}_{\epsilon_{\mu}}$ 
satisfy the (coupled) ordinary first order differential equations
\begin{align}
\left(\frac{d}{dx_{\mu}}+\frac{1}{4} \frac{X_{\mu}^{'}}{X_{\mu}}
+\frac{\epsilon_{\mu} Y_{\mu}}{X_{\mu}}
\right) \chi^{(\mu)}_{-\epsilon_{\mu}}+
\frac{(-1)^{\mu}(\epsilon_{\mu})^{n-\mu} Q(\epsilon_{\mu} x_{\mu})}
{\sqrt{(-1)^{\mu-1} X_{\mu}}}  \chi^{(\mu)}_{\epsilon_{\mu}}=0 ~.
\end{align}
The demonstrated separation is justified by the existence of the rank-$2$ closed CKY
tensor.  As in the four-dimensional case there exist symmetry operators 
which commute with the Dirac operator.
\cite{Benn:1997a,Benn:1997b,Kress:1997,Carliglia:2011}

\section{Classification of Higher-Dimensional Spacetimes}
In this section we present a classification of spacetimes admitting a rank-$2$ closed CKY tensor.
A key property of such spacetimes is that there is a family of commuting Killing tensors and Killing vectors.
In full generality, the classification is quite complicated. We will discuss this problem in section 5.2. 
We first describe a special class of spacetimes admitting a PCKY tensor.
The Kerr-NUT-(A)dS spacetimes are included in this class.
\subsection{Uniqueness of Kerr-NUT-(A)dS spacetime}

\begin{theorem} \label{uniqueness}
Let $(M,\bm{g})$ be a (2n+$\varepsilon$)-dimensional spacetime with a PCKY tensor. 
Then, the metric $\bm{g}$ is locally written as \eqref{KerrNUT1} for even dimensions or 
\eqref{KerrNUT2} for odd dimensions:
\begin{align} \label{KNDO}
\bm{g} = \sum_{\mu=1}^{n}\frac{U_\mu}{X_\mu} dx_{\mu}^2+
\sum_{\mu=1}^{n} \frac{X_{\mu}}{U_\mu} \left( \sum_{k=0}^{n-1} A_{\mu}^{(k)}
 d\psi_k \right)^2+\varepsilon \frac{c}{A^{(n)}}\left( \sum_{k=0}^{n} A^{(k)}
 d\psi_k \right)^2~~
\end{align}
with arbitrary functions $X_\mu$ of one variable $x_\mu$.
In particular, the Kerr-NUT-(A)dS spacetime is the only Einstein space admitting a PCKY tensor. 
\end{theorem}\quad

Theorem \ref{uniqueness} was first proved \cite{Houri:2007} by assuming that 
the vector field $\bm{\xi}=\xi^a \partial_a$ defined by Eq.\ \eqref{ckyD}
satisfies the following conditions
\begin{align} \label{LieD}
\pounds_{\xi}\,\bm{g}=0 ~,~~~ \pounds_{\xi}\,\bm{h}=0 ~. 
\end{align}
Afterward it was shown that both these conditions hold from the existence of the PCKY. \cite{Krtous:2008}\\

For simplicity, we sketch the proof restricted to even dimensions $\varepsilon=0$.
Recall that a PCKY tensor $\bm{h}$ is a rank-2 closed CKY tensor of maximal order $(n, n+\varepsilon)$. 
The $n$ eigenvalues of $\bm{h}$, denoted by $\{x_\mu \}_{\mu=1,\cdots,n}$, are functionally independent. 
Then, one can introduce an orthonormal basis in which
\begin{align}
\bm{g}= \sum_{\mu=1}^n(\bm{e}^\mu\bm{e}^\mu+\bm{e}^{\hat{\mu}} \bm{e}^{\hat{\mu}}) ~,~~~
\bm{h}= \sum_{\mu=1}^n x_\mu\,\bm{e}^\mu \wedge \bm{e}^{\hat{\mu}} ~.
\end{align}
We refer to $\{\bm{e}^\mu,\bm{e}^{\hat{\mu}}\}$ as the canonical basis associated with a PCKY tensor.
The basis is fixed up to two-dimensional rotations in each of 2-planes $\bm{e}^\mu\wedge\bm{e}^{\hat{\mu}}$.
This freedom allows us to choose the components of the vector field as $\{\xi^\mu=0,\,\xi^{\hat{\mu}} \ne 0 \}$,
i.e.,
\begin{align} \label{guzai}
\bm{\xi}=\sum_{\mu=1}^n \sqrt{Q_\mu}\,\bm{e}_{\hat{\mu}}
\end{align}
with the dual basis $\{\bm{e}_\mu,\bm{e}_{\hat{\mu}}\}$. 
Here, $Q_\mu~(\mu=1,\cdots,n)$ are arbitrary functions.
It is shown that the eigenvalues of $\bm{h}$ have orthogonal gradients:
\begin{align}
\bm{e}_\mu(x_\nu)=\sqrt{Q_\mu}\delta_{\mu \nu} ~,~~~ \bm{e}_{\hat{\mu}}(x_\nu)=0 ~.
\end{align}

The next step is to consider the integrability condition of the PCKY equations
\begin{align} \label{Eq3_0.2}
\nabla_a h_{bc}= \xi_c g_{ab}-\xi_b g_{ac}.
\end{align} 
These equations are overdetermined. 
By differentiating and skew symmetrizing the equations, we obtain
\begin{align}\label{integral1}
R_{abd}{}^fh_{fc}-R_{abc}{}^fh_{fd}
=g_{bc} \nabla_{a} \xi_d-g_{ac} \nabla_{b} \xi_d-g_{bd} \nabla_{a} \xi_c+g_{ad} \nabla_{b} \xi_c.
\end{align}
We shall use the canonical basis. If we restrict the indices to $c=\mu$ and $d=\hat{\mu}$, 
then the left-hand side of Eq.\ \eqref{integral1} identically vanishes by the property
of the Riemann curvature, $R_{ab\mu \mu}=R_{ab\hat{\mu} \hat{\mu}}=0$. 
Hence, Eq.\ \eqref{integral1} reduces to\footnote{
For simplicity, we have used the following notaions: 
$$\nabla_a\xi_\nu=(e_\nu)^b\nabla_a\xi_b ~,~~~ 
\nabla_\mu\xi_\nu=(e_\mu)^a(e_\nu)^b\nabla_a\xi_b ~,~~~ \text{etc} ~.$$
}
\begin{align}\label{integral2}
\delta_{b \mu} \nabla_{a} \xi_{\hat{\mu}}-\delta_{a \mu} \nabla_{b} \xi_{\hat{\mu}}-\delta_{b\hat{\mu}} \nabla_{a} \xi_\mu+\delta_{a\hat{\mu}} \nabla_{b} \xi_\mu=0.
\end{align}
The specialization of the indices yields:
\begin{eqnarray}\label{com}
& &
\nabla_\mu \xi_\nu=\nabla_\mu \xi_{\hat{\nu}}
=\nabla_{\hat{\mu}} \xi_{\nu}
=\nabla_{\hat{\mu}} \xi_{\hat{\nu}}
=0 ~,~~~(\mu \ne \nu)\nonumber\\
& &\nabla_{\mu} \xi_{\mu}+\nabla_{\hat{\mu}} \xi_{\hat{\mu}}=0.
\end{eqnarray}
Combining Eq.\ \eqref{Eq3_0.2} and Eq.\ \eqref{com} with the following identity
\begin{align}
\pounds_{\xi} h_{ab}=\xi^c \nabla_c h_{ab}+h_{cb} \nabla_a \xi_c+h_{ac} \nabla_b \xi_c,
\end{align}
we obtain $\pounds_{\xi} \bm{h}=0$.

Now let us consider the first structure equation
\begin{align}
d\bm{e}^a+\bm{\omega}^a_{~b} \wedge \bm{e}^b=0
\end{align}
corresponding to the canonical basis. The following lemma is obtained\cite{Houri:2009}:\\

\begin{lemma}
The connection 1-form $\omega_{ab}$ takes the form:
\begin{align}\label{connection}
\bm{\omega}_{\mu \nu}
=& -\frac{x_\nu \sqrt{Q_\nu}}{x_\mu^2-x_\nu^2}\bm{e}^\mu
   -\frac{x_\mu \sqrt{Q_\mu}}{x_\mu^2-x_\nu^2}\bm{e}^\nu~, \nonumber\\
\bm{\omega}_{\mu \hat{\nu}}
=& \frac{x_\mu \sqrt{Q_\nu}}{x_\mu^2-x_\nu^2}\bm{e}^{\hat{\mu}}
   -\frac{x_\mu \sqrt{Q_\mu}}{x_\mu^2-x_\nu^2}\bm{e}^{\hat{\nu}} ~,~~(\mu \ne \nu) \nonumber\\
\bm{\omega}_{\hat{\mu} \hat{\nu}}
=& -\frac{x_\mu \sqrt{Q_\nu}}{x_\mu^2-x_\nu^2}\bm{e}^\mu
   -\frac{x_\nu \sqrt{Q_\mu}}{x_\mu^2-x_\nu^2}\bm{e}^\nu~,
\end{align}
and for $\mu=1,2, \cdots ,n $(with no sum),
\begin{align}
\bm{\omega}_{\mu \hat{\mu}}
=& \sum_{\rho \ne \mu}\frac{x_\mu \sqrt{Q_\rho}}{x_\mu^2-x_\rho^2}\bm{e}^{\hat{\rho}}
   -\frac{1}{\sqrt{Q_\mu}}\sum_{\rho \ne \mu}\frac{x_\mu Q_\rho}{x_\mu^2-x_\rho^2}\bm{e}^{\hat{\mu}}\nonumber\\
+& \frac{1}{\sqrt{Q_\mu}}\sum_{\rho=1}^n (\nabla_\rho \xi_\mu\,\bm{e}^\rho
   +\nabla_{\hat{\rho}} \xi_\mu\,\bm{e}^{\hat{\rho}}).
\end{align}
\end{lemma}\quad

Using the connection $\bm{\omega}_{ab}$ and Eq.\ \eqref{guzai} 
one can evaluate the covariant derivation $\nabla_a \xi^b$. 
On the other hand we already know some components \eqref{com} from the integrability of the PCKY equations.
As a result we have several consistency conditions, which are summarized as
\begin{align}
\nabla_{\hat{\mu}} \xi_{\hat{\mu}}=e_{\hat{\mu}}(\sqrt{Q_\mu}),~~\nabla_{\mu} \xi_{\hat{\mu}}=e_{\mu} (\sqrt{Q_\mu})-\sum_{\rho \ne \mu} \frac{x_\mu Q_\rho}{x_\mu^2-x_\rho^2},
\end{align}
and
\begin{align}\label{QM}
e_{\hat{\mu}} (\sqrt{Q_\nu})=0,~~e_\mu (\sqrt{Q_\nu})+\frac{x_\mu \sqrt{Q_\mu} \sqrt{Q_\nu}}{x_\mu^2-x_\nu^2}=0
\end{align}
for $\mu \ne \nu$. 
Further non-trivial conditions are obtained from the Jacobi identity, 
$[[\bm{e}_a,\bm{e}_b],\bm{e}_c]+$(cyclic)$=0$. 
The commutators $[\bm{e}_a, \bm{e}_b]$ evaluated by the covariant derivation
give rise to the following conditions:
\begin{align}
\nabla_{\hat{\mu}} \xi_{\hat{\mu}}=0,~~\nabla_{\mu} \xi_{\hat{\mu}}+\nabla_{\hat{\mu}} \xi_{\mu}=0,
%0\nabla_{\hat{\mu}} \sqrt{Q_\mu},~~\nabla_{\mu} \xi_{\hat{\mu}}=\nabla_{\mu} \sqrt{Q_\mu}-\sum_{\rho \ne \mu} \frac{x_\mu Q_\rho}{x_\mu^2-x_\rho^2},
\end{align}
Thus we have seen that the vector field $\xi$ satisfies $\nabla_{(a} \xi_{b)}=0$ for all components.
It turns out that $\bm{\xi}$ is a Killing vector obeying the equation $\pounds_{\xi} \bm{g}=0$.\footnote{
This condition can be easily proved if the Einstein condition is imposed for the metric,
because it is shown from (\ref{integral1}) that
\begin{align}
\nabla_{(a} \xi_{b)}=\frac{1}{4(n-1)}(h_a^{~c}R_{cb}+h_b^{~c}R_{ca}).
\end{align}
The Ricci tensor $R_{ab}$ is proportional to the metric in the Einstein spaces, 
so that we immediately obtain $\nabla_{(a} \xi_{b)}=0$ from the equation above. 
This result was first demonstrated by Tachibana\cite{Tachibana:1969}. }

As the final step, we introduce a new basis $\{ \bm{v}_\mu, \bm{\eta}^{(j)} \}~(\mu=1,\cdots,n; j=0,\cdots, n-1)$,
\begin{align}
\bm{v}_\mu=\frac{\bm{e}_\mu}{\sqrt{Q_\mu}} ~,~~~
\bm{\eta}^{(j)}=\sum_{\mu=1}^n A^{(j)}_\mu \sqrt{Q_\mu} \bm{e}_{\hat{\mu}},
\end{align}
where $A^{(j)}_\mu$ is again given by Eq.\ \eqref{yu} and $\bm{\eta}^{(0)} \equiv \bm{\xi}$. 
These vector fields $v_\mu$ and $\eta^{(j)}$ geometrically represent eigenvectors of Killing tensors 
and Killing vectors, respectively. 
One can easily show that
\begin{align}
[\bm{v}_\mu, \bm{v}_\nu]=0,
\end{align}
which implies that there are local coordinates $x_\mu~(\mu=1, \cdots ,n)$ 
such that $\bm{v}_\mu=\partial/\partial x_\mu$. Then the functions $Q_\mu$ take the form
\begin{align}
Q_\mu=\frac{X_\mu}{U_\mu},~~U_{\mu} = \prod_{\substack{\nu=1\\(\nu \neq \mu)}}^n
( x_{\mu}^2 - x_{\nu}^2 ),
\end{align}
where each $X_\mu$ is a function depending on $x_\mu$ only. 
This can be obtained easily using the differential equation \eqref{QM} together with
\begin{align}
e_{\hat{\mu}}(\sqrt{Q_\mu})=\nabla_{\hat{\mu}}\xi_{\hat{\mu}}=0.
\end{align}
Now, we can also prove the commutativity of the remaining vector fields $\bm{\eta}^{(j)}$, 
which is essentially the same as proposition \ref{Prop6OfCKY}\,:
\begin{align}
[\bm{v}_\mu, \bm{\eta}^{(j)}]=[\bm{\eta}^{(i)}, \bm{\eta}^{(j)}]=0.
\end{align}
This introduces the local coordinates $\psi_j~(j=0,\cdots,n-1)$ 
such that $\bm{\eta}^{(j)}=\partial/\partial \psi_j$. Finally, we have
\begin{align}
\frac{\partial}{\partial x_\mu}=\sqrt{\frac{U_\mu}{X_\mu}}\,\bm{e}_\mu ~,~~~
\frac{\partial}{\partial \psi_j}=\sum_{\mu=1}^n \sqrt{\frac{X_\mu}{U_\mu}}A^{(j)}_\mu\,\bm{e}_{\hat{\mu}},
\end{align}
which reproduces the orthonormal basis \eqref{ortho1} , and hence the required metric \eqref{KNDO}.

\subsection{Generalized Kerr-NUT-(A)dS spacetime}
Let $(M, \bm{g})$ be a $D$-dimensional spacetime 
with a rank-$2$ closed CKY tensor $\bm{h}$. 
When $\bm{h}$ is a PCKY tensor, it has functionally independent $n$ eigenvalues.
For the general $\bm{h}$, it is important to know how many of the eigenvalues are functionally independent. 
This information tells us the number of independent Killing tensors and Killing vectors.
To do so we introduce a rank-2 conformal Killing tensor $K_{ab}=h_{ac} h_b{}^c$ 
associated with $\bm{h}$ according to proposition \ref{Prop1OfCKY}. 
Since this tensor is symmetric, $K^{a}{}_{b}$ can be diagonalized at any point on
$M$. Let
$x_{\mu}^2$ ($\mu=1,\cdots, \ell$) 
and 
$a_i^2$ 
($i=1,\cdots, N$) be
the non-constant eigenvalues and the non-zero constant eigenvalues of  $K^a{}_b$, respectively. 
Taking account of the multiplicity we write the eigenvalues as
\begin{align}\label{eigen}
\{
\underbrace{x_1^2, \dotsm,  x_1^2}_{2n_1},
\dotsm, 
\underbrace{x_\ell^2, \dotsm, x_\ell^2}_{2n_n}, 
\underbrace{a_1^2, \dotsm, a_1^2}_{2m_1},
\dotsm,
\underbrace{a_N^2, \dotsm, a_N^2}_{2m_N}, 
\underbrace{0,\dotsc, 0}_{m_0} \}.
\end{align}
The total number of the eigenvalues is equal to the spacetime dimension:
$D=2 (|n| + |m| )+ m_0$.
Here $|n| = \sum_{\mu=1}^\ell n_{\mu}$ , $|m|=\sum_{i=1}^N m_i$ and $m_0$ represents the number of zero eigenvalues. \\

\begin{lemma}
The multiplicity constant $n_\mu$ of the non-constant eigenvalues $x_\mu^2$ is equal to one\cite{Houri:2009}.
\end{lemma}\quad

Independent Killing tensors and Killing vectors are relevant to non-constant eigenvalues $x_\mu^2$.
Indeed the general construction discussed in section 3.2 yields 
that the Killing tensors $\bm{K}^{(j)}~(j=0,\cdots,\ell-1)$ and 
the Killing vectors $\bm{\eta}^{(j)}~(j=0,\cdots,\ell-1+\delta)$ are independent quantities, 
i.e. the order of the CKY tensor is $(\ell, \ell+\delta)$ with $\delta=0$ for $m_0 > 1$ and $\delta=1$ for $m_0=1$.
The construction of the metric is rather parallel to 
that of the PCKY case except for consideration of constant eigenvalues. 
Associated with the non-zero constant eigenvalues the spacetime admits K\"ahler
submanifolds of the same dimensions as the multiplicity of them, 
and the metric becomes the ``Kaluza-Klein metric" on 
the bundle over the K\"ahler manifolds whose fibers are given by theorem \ref{uniqueness}.
More precisely we prove the following classification\cite{Houri:2009}.\\

\begin{theorem} \label{theorem:5.3}
Let $(M,\bm{g})$ be a $D$-dimensional spacetime with a rank-2 closed CKY tensor 
with order $(\ell, \ell+\delta)$. 
Then the metric $g$ takes the forms
\begin{align} \label{gCKY}
\bm{g} = 
\sum_{\mu=1}^{\ell} \frac{U_\mu}{X_\mu}dx_\mu^2+
\sum_{\mu=1}^{\ell} \frac{X_\mu}{U_\mu} \left( \sum_{k=0}^{\ell-1} A^{(k)}_\mu
\bm{\theta}_k \right)^2
+ \sum_{i=1}^{N}\prod_{\mu=1}^{\ell}(x_{\mu}^2- a_i^2)\bm{g}^{(i)}+
A^{(\ell)} \, \bm{g}^{(0)} ~,
\end{align}
where $\bm{g}^{(i)}$ are 
K\"ahler metrics on $2 m_i$-dimensional K\"ahler manifolds $B^{(i)}$.
The metric $\bm{g}^{(0)}$ is, in general, any
 metric on an $m_0$-dimensional manifold $B^{(0)}$ associated with the zero eigenvalues, but if $m_0=1$, 
$\bm{g}^{(0)}$ can take the
special form:
\begin{align}\label{g0sp}
\bm{g}^{(0)}_{\mathrm{special}} = \frac{c}{(A^{(\ell)})^2} \left( \sum_{k=0}^{\ell} A^{(k)}
\bm{\theta}_k \right)^2
\end{align}
with a constant $c$. The functions $U_{\mu}$, $A^{(k)}$ and $A^{(k)}_\mu$ 
are defined by \eqref{yu}, and $X_\mu$
is a function depending on $x_{\mu}$ only.
The 1-forms $\theta_k$ satisfy
\begin{align}\label{theta}
d\bm{\theta}_k+ 2\sum_{i=1}^N (-1)^{\ell-k} a_i^{2\ell-2k-1} \bm{\omega}^{(i)}=0, 
%\qquad
%k=0,1,\dotsc, \ell-1+\varepsilon,
\end{align}
where $\bm{\omega}^{(i)}$ represents the K\"ahler form on $B^{(i)}$.
\end{theorem}\quad

The spacetime $M$ has the bundle structure: 
the base space is an $(m_0+2|m|)$-dimensional product space $B^{(0)} \times B^{(i)} \times \cdots \times B^{(N)}$ 
of the general manifold $B^{(0)}$ and the K\"ahlar manifolds $B^{(i)}~(i=1,\cdots,N)$, 
while the fiber spaces are $2\ell$-dimensional spaces with the metric \eqref{KNDO}.
%We call the Kaluza-Klein metric \eqref{gCKY} the generalized Kerr-NUT-(A)dS metric. 
The fiber metric in theorem 5.1
is twisted by the K\"ahler form $\bm{\omega}^{(i)}$; the $1$-form $d\psi_k$ in \eqref{KNDO} 
is replaced by the $1$-form
$\bm{\theta}_k$. The K\"ahler form is  locally written as $\bm{\omega}^{(i)} = d\bm{\beta}^{(i)}$, 
and so \eqref{theta} is equivalent to
\begin{align} 
\bm{\theta}_k = d \psi_k - 2 \sum_{i=1}^N (-1)^{\ell-k} a_i^{2\ell-2k-1} \bm{\beta}^{(i)}.
\end{align}
If we use the 1-form $\bm{\beta}^{(i)}$, then
the CKY tensor 
can be written in a manifestly closed form:
\begin{align}
\bm{h} = d \Biggl( \frac{1}{2} \sum_{k=0}^{\ell-1} A^{(k)} d \psi_k
+ \sum_{i=1}^N a_i \prod_{\mu=1}^\ell ( x_{\mu}^2- a_i^2  ) \bm{\beta}^{(i)} \Biggr).
\end{align}
In order to see the eigenvalues \eqref{eigen} it is convenient to introduce the orthonormal basis like \eqref{ortho1} and 
\begin{align}
\bm{e}^{\alpha}_{(i)}=\sqrt{\prod_{\mu}(x_\mu^2-a_i^2)}\,\tilde{\bm{e}}^{\alpha}_{(i)} ~,~~~ 
\bm{e}^{\hat{\alpha}}_{(i)}=\sqrt{\prod_{\mu}(x_\mu^2-a_i^2)}\tilde{\bm{e}}^{\hat{\alpha}}_{(i)} ~,
\end{align}
where
\begin{align}
\bm{g}^{(i)}=\sum_{\alpha=1}^{m_i}\bigl(\tilde{\bm{e}}^{\alpha}_{(i)} \tilde{\bm{e}}^{\alpha}_{(i)}
+\tilde{\bm{e}}^{\hat{\alpha}}_{(i)} \tilde{\bm{e}}^{\hat{\alpha}}_{(i)}\bigr)~,~~~\bm{\omega}^{(i)}=
\sum_{\alpha=1}^{m_i}\tilde{\bm{e}}^{\alpha}_{(i)} \wedge \tilde{\bm{e}}^{\hat{\alpha}}_{(i)}.
\end{align}
Then we have
\begin{align}\label{hCKY}
\bm{h}=\sum_{\mu=1}^\ell x_\mu \bm{e}^\mu \wedge \bm{e}^{\hat{\mu}}
+\sum_{i=1}^N \sum_{\alpha=1}^{m_i} a_i \bm{e}^{\alpha}_{(i)} \wedge  \bm{e}^{\hat{\alpha}}_{(i)},
\end{align}
where the coefficients $\{ x_\mu, a_i \}$ represent the eigenvalues.

The total metric $\bm{g}$ includes arbitrary functions $X_\mu=X_\mu(x_\mu)$ of the single coordinate $x_\mu$. These are fixed if we impose
the Einstein equations for the metric, $R_{ab}=\Lambda g_{ab}$.\\

%%%%%%%%%%%%%%%%%%%%%%%%%%%%%%%%%%%%%%%%%%%%%%%%%%%%%
\begin{theorem} \label{theorem:5.4}
The metric (\ref{gCKY}) is an Einstein metric if and only if the following
conditions hold\cite{Houri:2008b}:
\begin{itemize}
\item [(i)]$X_{\mu}$ takes the form
\begin{align} \label{EC}
X_{\mu}
=\frac{1}{ (x_{\mu})^{m_0-1} \prod_{i=1}^N (x_\mu^2-a_i^2)^{m_i}}
\Biggl( d_{\mu}
+ \int \mathcal{X}(x_{\mu}) \, x_{\mu}^{m_0-2} \prod_{i=1}^N( x_{\mu}^2 - a_i^2)^{m_i} \, d x_{\mu}
\Biggr),
\end{align}
where
\begin{align}\label{kai}
\mathcal{X}(x) = \sum_{i=0}^\ell \alpha_i x^{2i}, \qquad
\alpha_\ell= - \Lambda.
\end{align}
For the special case \eqref{g0sp} $\mathcal{X}(x)$ is replaced by
\begin{align}
\mathcal{X}_{\mathrm{special}} (x) = \frac{\alpha_{-1}}{x}+\sum_{i=0}^\ell \alpha_i x^{2i}
\end{align}
with
\begin{align}
\alpha_0=(-1)^{\ell-1} 2c \sum_{j=1}^N \frac{m_j}{a_j^2}, \qquad
\alpha_{-1} = (-1)^{\ell-1} 2c.
\end{align}
Here $\{ \alpha_k \}_{k=1,2,\cdots, \ell-1}$ and $\{d_{\mu} \}_{\mu=1,2, \cdots, \ell}$ are free parameters.
(In \eqref{kai}  $\alpha_0$ is also a free parameter.)
\item [(ii)] $\bm{g}^{(i)}(i=1,\cdots,N)$ are $2m_i$-dimensional K\"ahler-Einstein metrics with the cosmological constants
\begin{align}
 \lambda^{(i)}=(-1)^{\ell-1}\mathcal{X}(a_i).
\end{align}
\item [(iii)] $\bm{g}^{(0)}$ is an $m_0$-dimensional Einstein metric with the cosmological constant
\begin{align}
 \lambda^{(0)}=(-1)^{\ell-1}\alpha_0.
\end{align}
\end{itemize}
\end{theorem}\quad

Theorems \ref{theorem:5.3} and \ref{theorem:5.4} give a complete local classification of Einstein spacetimes admitting a rank-$2$ closed CKY tensor.
We call these metrics the generalized Kerr-NUT-(A)dS metrics. Important examples are given by a special class of the Kerr-(A)dS metrics.
The general $D$-dimensional Kerr-(A)dS metric has an isometry ${\bf{R}} \times U(1)^n,~n=[(D-1)/2]$. 
\cite{Gibbons:2004a,Gibbons:2005}
This symmetry is enhanced when some of rotation parameters coincide.
Such metrics can be written as the generalized Kerr-NUT-(A)dS metrics with the Fubini-Study metrics on the base space 
$B \equiv \mathbb{CP}^{m_1-1} \times \cdots \times \mathbb{CP}^{m_N-1} $($m_i$ represents the multiplicity of the rotation parameters.).
In particular the $(D=2n+1)$-dimensional metric with all rotation parameters equal has an isomery ${\bf{R}} \times U(n)$ and $B= \mathbb{CP}^{n-1}$.  

The generalized Kerr-NUT-(A)dS metrics are also
interesting
from the point of view of AdS/CFT correspondence. 
Indeed, BPS limit leads odd dimensional Einstein metrics to
Sasaki-Einstein metrics\cite{Hashimoto:2004,Cvetic:2005a,Cvetic:2005b,Chen:2006b}
and even dimensional Einstein metrics to
Calabi-Yau metrics\cite{Oota:2006,Lu:2007,Martelli:2009}. Especially,
the five-dimensional Sasaki-Einstein
metrics have emerged quite naturally in the 
AdS/CFT correspondence\cite{Gauntlett:2004,Klebanov:1998}. 
The related topics will be briefly discussed in section 6.2.

%%%%%%%%%%%%%%%%%%%%%%%%%%%%%%%%%%%%%%%%%%%%%%%%%%
%%%%%%%%%%%%%%%%%%%%%%%%%%%%%%%%%%%%%%%%%%%%%%%%%%
%%%%%%%%%%%%%%%%%%%%%%%%%%%%%%%%%%%%%%%%%%%%%%%%%%
\section{Further Developments}

\subsection{Killing-Yano symmetries in the presence of skew-symmetric torsion}
In this section we discuss the symmetries of black holes of more general theories 
with additional matter content, such as various supergravity theories or string theories.
These black holes are usually much more complicated and the presence
of matter tends to spoil many of the elegant properties of the Kerr black hole.
Recently, there has been success in constructing charged rotating black hole solutions 
of the supergravity theories\cite{Cvetic1:1996,Chong:2005a,Cvetic2:1996,Cvetic3:1996,Chow2:2010}.
This is because these theories possess global symmetries, 
and they provide a generating technique that produces charged solutions
from asymptotically flat uncharged vacuum solutions. 
However, it is known that such a generating technique 
does not work for search of AdS black hole solutions of gauged supergravity theories. In these theories
some guesswork is required rather than systematic construction.
\cite{Chong:2005a,Chong:2005b,Chong:2005c,Chong:2005d,Gauntlett:2003,Kunduri:2006,Chong:2007,
Mei:2007,Chow:2008,Chow1:2010,Chow3:2010,Chow4:2010}

Here, we discuss a Killing-Yano symmetry in the presence of skew-symmetric torsion.
The spacetimes with skew-symmetric torsion occur naturally in supergravity theories, 
where the torsion may be identified with a
$3$-form field strength. \cite{Strominger:1986} Black hole spacetimes of 
such theories are natural candidates to admit the Killing-Yano symmetry with torsion.
This generalized
symmetry
was first introduced by Bochner and Yano
\cite{Yano:1953} from the mathematical point of view and
recently rediscovered \cite{Wu:2009a,Kubiznak:2009a,Wu:2009b,Ahmedov:2009} as a hidden
symmetry of the Chong--Cvetic--L\"u--Pope rotating black hole of $D=5$
minimal gauged supergravity \cite{Chong:2005b}.
Furthermore, this was found in the Kerr-Sen black hole solution of effective string theory 
\cite{Hioki:2008,Chow2:2010}
and its higher-dimensional generalizations\cite{Houri:2010b,Kubiznak:2011}.
The discovered generalized symmetry shears almost identical properties with its vacuum cousin; 
it gives rise to separability of the Hamilton-Jacobi, 
Klein-Gordon and Dirac equations in these backgrounds.

These results produce the natural question of whether there are
some other physically interesting spacetimes which admit 
the Killing-Yano symmetry with skew-symmetric torsion.
It is the purpose of this section to present a family of spacetimes
admitting the generalized symmetry with torsion, and hence to show that such symmetry
is more widely applicable.

\subsubsection{Generalized Killing-Yano symmetries}
We first recall some notations concerning a connection with totally skew-symmetric torsion.
Let $\bm{T}$ be a 3-form and $\nabla^T$ be a connection defined by
\begin{align}
\nabla^T_X \bm{Y} = \nabla_X \bm{Y} + \frac{1}{2} \sum_aT(X,Y,e_a)\bm{e}_a~,
\end{align}
where 
$\nabla_a$ is the Levi-Civita connection and $\{\bm{e}_a\}$ is an orthonormal basis. 
One can characterize this connection geometrically: the connection $\nabla^T_a$ satisfies a metricity condition $\nabla^T_a g_{bc}=0$, 
and geodesic-preserving if and only if the torsion $\bm{T}$ lies in $3$-form. The second condition means that
the connection $\nabla^T_a$ has the same geodesic as $\nabla_a$.
For a $p$-form $\bm{\Psi}$ the covariant derivative is calculated as
\begin{align}
\nabla^T_X\bm{\Psi} 
= \nabla_X\bm{\Psi}-\frac{1}{2}\sum_a(\bm{X}\hook \bm{e}_a\hook \bm{T})\wedge (\bm{e}_a\hook\bm{\Psi}) ~.
\end{align} 
Then, we define the differential operators
\begin{align}
d^T\bm{\Psi} = \sum_a \bm{e}^a \wedge\nabla^T_{e_a}\Psi ~,~~~
\delta^T\bm{\Psi} = -\sum_a \bm{e}_a\hook \nabla^T_{e_a}\bm{\Psi} ~.
\end{align}

A {\it generalized conformal Killing-Yano (GCKY) tensor}
$\bm{k}$ was introduced \cite{Kubiznak:2009a} as  a $p$-form satisfying for any vector field $\bm{X}$
\begin{align}
\nabla^T_X\bm{k} 
= \frac{1}{p+1}\bm{X}\hook d^T\bm{k} - \frac{1}{D-p+1}\bm{X}^*\wedge\delta^T\bm{k} ~. \label{1-5}
\end{align}
In analogy with Killing-Yano tensor with respect to the Levi-Civita connection, 
a GCKY $p$-form $\bm{f}$ obeying $\delta^T\bm{f}=0$ is called
a {\it generalized Killing-Yano (GKY) tensor}, 
and a GCKY $p$-form $\bm{h}$ obeying $d^T\bm{h}=0$ is called 
a {\it generalized closed conformal Killing-Yano (GCCKY) tensor}. \\

\begin{prop} \label{bpGCKY}
GCKY tensors possess the following basic properties\cite{Kubiznak:2009a}:\\[-0.3cm]
\begin{enumerate}
\item A GCKY 1-form is equal to a conformal Killing 1-form.\\[-0.3cm]
\item The Hodge star $*$ maps GCKY $p$-forms into GCKY ($D-p$)-forms.
In particular, the Hodge star of a GCCKY $p$-form is a GKY ($D-p$)-form and vice versa.\\[-0.3cm]
\item When $\bm{h}_1$ and $\bm{h}_2$ is a GCCKY $p$-form and $q$-form, 
then $\bm{h}_3=h_1\wedge \bm{h}_2$ is a GCCKY ($p+q$)-form.\\[-0.3cm]
\item Let $\bm{k}$ be a GCKY $p$-form. Then
\begin{align}
Q_{ab} = k_{ac_1\cdots c_{p-1}}k_b{}^{c_1\cdots c_{p-1}}
\end{align}
is a rank-2 conformal Killing tensor.
In particular, $\bm{Q}$ is a rank-2 Killing tensor if $\bm{k}$ is a GKY tensor.
\end{enumerate}
\end{prop}\quad

From these properties, we find that a GCCKY tensor also generates the tower of commuting Killing tensors 
in the similar way to section 3.2.
On the other hand, there is a difference between the closed CKY 2-form and the GCCKY 2-form.
\cite{Houri:2010b} 
%In the presence of torsion,
%Killing tensors commute in the following sense instead of the proposition 3.4:
%\begin{align}
%[K^{(j)},K^{(\ell)}]^T_{abc}
%\equiv K^{(j)}_{e(a}\nabla^{Te}K^{(\ell)}_{bc)}-K^{(\ell)}_{ea}\nabla^{Te}K^{(j)}_{bc)} = 0 ~.
%\end{align}
%This means that the integrals of motion generated from Killing tensors don't commute 
%with respect to Poisson bracket because the term depending on the torsion appears.
%We further see such a torsion anomaly in the construction of Killing vectors.
When the torsion is present,
neither $\delta^T\bm{h}$ nor
$\delta \bm{h}$ are in general Killing vectors and the whole construction in section 3.2 breaks down.
In this way, torsion anomalies appear everywhere 
in considering geometry with the GCCKY 2-form.
Does the existence of a GCCKY 2-form $\bm{h}$ imply the existence of the isometries?
The Kerr--Sen black hole spacetime (and more generally the charged Kerr-NUT metrics) 
studied in the next section provides an example of geometries with a non-degenerate 
GCCKY 2-form and $n+\varepsilon$ isometries.

We should emphasize that torsion anomalies appear in
contributions of a GCCKY 2-form to separation of variables in field equations.
As already explained, separation of variables in differential equations
is deeply related to the existence of symmetry operators,
which commute between themselves and whose number is that of dimensions.
In the presence of torsion,
the commutator between a symmetry operator generated by a Killing tensor
and the laplacian doesn't vanish in general.
This means that a GCCKY 2-form no longer generates symmetry operators
for Klein-Gordon equation.
Similarly, it is known that the non-degenerate GCCKY 2-form doesn't 
in general generate symmetry operators for Dirac equation\cite{Houri:2010a},
while it is possible for primary CKY tensor.

\subsubsection{Charged rotating black holes with a GCCKY 2-form}
We have seen that, when the torsion is an arbitrary 3-form, 
one obtains various torsion anomalies and the implications 
of the existence of the generalized Killing-Yano symmetry are relatively weak 
compared with ordinary Killing-Yano symmetry. 
However, in the spacetimes where there is a natural 3-form
obeying the appropriate field equations, 
these anomalies disappear and the concept of generalized Killing-Yano symmetry 
may become very powerful.

Let us consider $D$-dimensional spacetimes admitting a GCCKY 2-form.
The GCKY equation is rather analogous to the CKY equation, which
%One may expect the metric on such spacetimes takes the form analogous to the theorem 4.4 
leads us to a fairly tight ansatz for the metric. Actually we consider the following metric:
\begin{align} \label{TCKY}
\bm{g} =& 
\sum_{\mu=1}^{\ell} \frac{U_\mu}{X_\mu}dx_\mu^2+
\sum_{\mu=1}^{\ell} \frac{X_\mu}{U_\mu} \left( \sum_{k=0}^{\ell-1} A^{(k)}_\mu
\bm{\theta}_k-\frac{1}{\Phi}\sum_{\nu=1}^{\ell}\frac{Y_\nu}{U_\nu}\sum_{k=0}^{\ell-1} A^{(k)}_\nu \bm{\theta}_k\right)^2 \nonumber\\
&+ \sum_{i=1}^{N}\prod_{\mu=1}^{\ell}(x_{\mu}^2- a_i^2)\bm{g}^{(i)}+
A^{(\ell)} \, \bm{g}^{(0)}.
\end{align}
The conventions are the same ones as Eq.\ \eqref{gCKY}. The only difference is in the second term, where
new functions $Y_\mu~(\mu=1,\cdots,\ell)$ are introduced. 
The functions $Y_\mu$ depend on the single variable $x_\mu$ like $X_\mu$ and $\Phi$ is defined by
\begin{align}
\Phi=1+\sum_{\mu=1}^{\ell}\frac{Y_\mu}{U_\mu}.
\end{align}
When we assume the following torsion 3-form
\begin{align}
\bm{T}
= -\sum_{\mu=1}^{\ell} \sqrt{\frac{X_\mu}{U_\mu}}\,\bm{e}^{\hat{\mu}}\wedge
  \Bigg( \sum_{\rho=1}^n \frac{\partial_\rho \Phi}{\Phi}
  \,\bm{e}^\rho \wedge\bm{e}^{\hat{\rho}}
  +\sum_{i=1}^{N} \Xi_i \sum_{\alpha=1}^{m_i} 
  \,\bm{e}^{\alpha}_{(i)}\wedge\bm{e}^{\hat{\alpha}}_{(i)}\Bigg),
\end{align}
where
\begin{align}
\Xi_i=\frac{2}{\Phi} \sum_{\mu=1}^\ell \frac{Y_\mu}{U_\mu} \frac{a_i}{x_\mu^2-a_i^2},
\end{align}
then there exists a rank-2 GCCKY tensor $\bm{h}$ which takes the form \eqref{hCKY}.

In supergravity theories, the metric $\bm{g}$ and the 3-form field strength 
$\bm{H}=d\bm{B}-\bm{A}\wedge d\bm{A}$ identified with the torsion $\bm{T}$
are required to satisfy the equations of motion which are generalization of the Einstein equations.
For this, in addition, a dilaton field $\phi$, and a Maxwell field $\bm{F}=d\bm{A}$ (2-form) are introduced, 
and the equations of
motion (in the string frame) can be written as
\cite{Sen:1992,Chow2:2010,Cvetic3:1996}
\begin{align}
& R_{ab}-\nabla_a \nabla_b \phi-F_{a}^{~c}F_{bc}-\frac{1}{4}H_{a}^{~cd}H_{bcd}=0 ~, \nonumber\\
%& G_{ab}= 
%  \sqrt{\frac{D-2}{2}}\nabla_a\nabla_b\varphi 
%  -\sqrt{\frac{D-2}{2}}g_{ab}\nabla^2\varphi-\frac{1}{2}\frac{D-2}{2}g_{ab}(\nabla\varphi)^2 \nonumber\\
%& \hspace{1.0cm}
%  +\Big(F_{(2)}{}_a{}^cF_{(2)}{}_{bc}-\frac{1}{2}g_{ab}||F_{(2)}||^2\Big)
%  +\frac{1}{4}\Big(H_{(3)}{}_a{}^{cd} H_{(3)}{}_{bcd}-g_{ab}||H_{(3)}||^2\Big) ~, \nonumber\\
& d\Big(e^{\phi} *\bm{F}\Big)=e^{\phi} *\bm{H}\wedge \bm{F} ~,~~~
  d\Big(e^{\phi} *\bm{H}\Big)=0 ~, \nonumber\\
& (\nabla\phi)^2+2 \nabla^2\phi+\frac{1}{2}F_{ab}F^{ab}+\frac{1}{12}H_{abc}H^{abc}-R = 0 ~.
\end{align}
These equations determine the unknown functions $Y_\mu$ as
\begin{align} \label{sol1}
Y_\mu=a X_\mu+\prod_{i=1}^N(x_\mu^2-a_i^2)(b_{\ell-1-N}x_\mu^{2(\ell-N-1)}+ \cdots + b_1 x_\mu^2+b_0) ~.
\end{align}
Then, the Maxwell potential $\bm{A}$ and the dilaton field $\phi$ become
\begin{align} \label{sol2}
\bm{A}=\frac{\kappa}{\Phi}\sum_{\mu=1}^\ell \frac{Y_\mu}{U_\mu} \sum_{k=0}^{\ell-1}A_{\mu}^{(k)} \bm{\theta}_k ~,~~~
\phi=\log \Phi.
\end{align}
In the expressions \eqref{sol1} and \eqref{sol2} the function $X_\mu$ is given by Eq.\ \eqref{EC} with $\Lambda=0$,
and $a$, $\kappa$, $\{b_\alpha \}_{\alpha=0, \cdots ,\ell-2-N}~(b_{\ell-1-N} \equiv a \kappa^2-1)$ 
are arbitrary constants with the range $ 0 \le N \le \ell-1$. 
When we take the special choices of the constants, 
the solutions represent charged rotating black hole solutions including the Kerr-Sen black hole 
and its higher-dimensional generalizations.
The torsion anomalies vanish on these black hole spacetimes, 
and hence one can expect integrable structure in various field equations like the Kerr background.

\subsection{Compact Einstein manifold}
In section 5 we have given an explicit local classification of
all Einstein metrics with a rank-2 closed CKY tensor. 
Remarkably, this class of metrics includes the Kerr-NUT-(A)dS
metrics, which are the most general Einstein metrics representing the rotating
black holes with spherical horizon. This section is concerned with the construction
of compact Einstein manifolds admitting the CKY tensor. This is an important issue to study the compactifications of higher-dimensional theories such as supergravity and superstring theories.
Examples of compact Einstein manifolds are rather rare. 
The first non-homogeneous example is an Einstein metric on the connected sum 
$\mathbb{CP}^{2} \sharp \overline{\mathbb{CP}^{2}}$.
This was discovered by Page\cite{Page:1979} as a certain limit of the 4-dimensional Kerr-de Sitter
black hole. Then, B\'erard-Bergery\cite{Berard:1982} and Page-Pope\cite{Page:1987} generalized Page's example. They 
independently obtained
Einstein metrics on the total space of $S^2$-bundles 
over K\"ahler-Einstein manifolds with
positive first Chern class. Furthermore, these metrics were generalized to
Einstein metrics on $S^2$-bundles with the base space of a product of 
K\"ahler-Einstein manifolds\cite{Wang:1998}.
As a different generalization, an infinite series of Einstein
metrics was constructed on $S^3$-bundles over $S^2$\cite{Hashimoto:2005}. They appear as a limit of
the 5-dimensional Kerr-de Sitter black hole. This work was generalized in the paper\cite{Gibbons:2005}
, where Einstein metrics were constructed on $S^n$-bundles over $S^2$ ($n \ge 2$).

The geometry with CKY tensor may be related to the K\"ahler geometry studied by Apostolov et.al
in a series of papers\cite{Apos:2004,Apos:2006}. They introduced the notion of a hamiltonian
2-form, and obtained a classification of all K\"ahler metrics admitting such a tensor. 
By taking a BPS limit, one can obtain such K\"ahler metrics from
the generalized Kerr-NUT-(A)dS metrics. 
Along this line several Sasaki-Einstein metrics 
and Calabi-Yau metrics were constructed. 
\cite{Hashimoto:2004,Cvetic:2005a,Cvetic:2005b,Oota:2006,Lu:2007,Martelli:2009,Kubiznak:2009,Chen:2006b}.

Finally, we briefly discuss the Einstein metrics over compact Riemannian manifolds
that are obtained from the metric \eqref{gCKY}. 
For general values of the parameters in \eqref{EC}
the metrics do not extend smoothly onto compact manifolds. However, this can be achieved for special choices of the parameters.
For simplicity we consider $N=1$ case: let $(B,g,\omega)$ be a 2$m$-dimensional
compact K\"ahler-Einstein manifolds with positive first Chern class $c_1(B)$.
One can write as $c_1(B)=p \alpha$, where $\alpha$ is an indivisible class in $H^2(B;\mathbb{Z})$ and $p$ is
a positive integer\footnote{The integer $p$ is always smaller than $m+1$ with equality only if $B$ is the complete projective space $\mathbb{CP}^m$.} . Let $P_{k_1,k_2,\cdots,k_n}$ be an $n$-torus bundle over $B$ classified by integers $(k_1,k_2,\cdots,k_n)$
and let $M^{(\varepsilon)}_{k_1,k_2,\cdots,k_n}~(\varepsilon=0,1)$ be the $S^{2n-\varepsilon}$-bundle 
over $B$ associated with $P_{k_1,k_2,\cdots,k_n}$.
Then we obtain the following theorems\cite{Oota:2011}:\\

\begin{theorem} \label{theorem:6.1}
If $k_\alpha$ are positive integers satisfying $0<k_1+k_2+\cdots+k_n<p$, 
then $M^{(0)}_{k_1,k_2,\cdots,k_n}$ admits an Einstein metric with positive scalar curvature.
\end{theorem}\quad

\begin{theorem} \label{theorem:6.2}
If $k_\alpha$ are positive integers, then $M^{(1)}_{k_1,k_2,\cdots,k_n}$ 
admits an Einstein metric with positive scalar curvature. 
In particular, if $k_1+k_2+\cdots+k_n=p$,
then $M^{(1)}_{k_1,k_2,\cdots,k_n}$ admits a Sasaki-Einstein metric.
\end{theorem}\quad

\begin{figure}[t]
\begin{center}
\includegraphics[width=10cm]{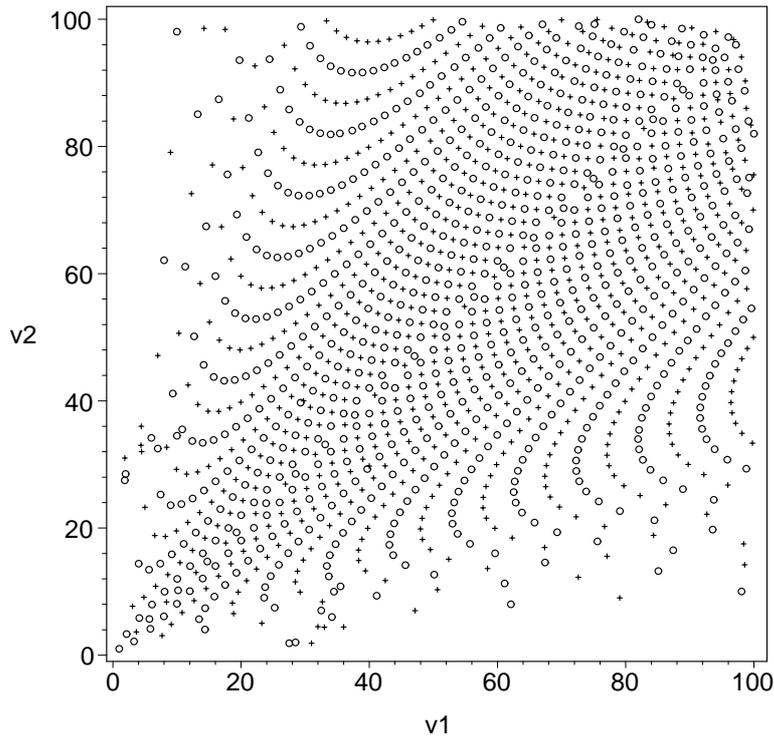}
\caption{Moduli space of Einstein metrics. We denote by circles and crosses the solutions to \eqref{intcond} for positive integers $k_1$ and $k_2$.  Circles have the topology of the non-trivial $S^3$-bundle over $S^2$.
Crosses correspond to topology $S^3 \times S^2$.}
\label{fig:moduli}
\end{center}
\end{figure}

These provide a unifying framework for the works 
\cite{Page:1979,Berard:1982,Page:1987,Lu:2004,Gibbons:2004,Hashimoto:2005,Gibbons:2005}, 
and at the same time
gives a new class of compact Einstein manifolds.
For example, we can obtain 5-dimensional Einstein metrics on $M^{(1)}_{k_1k_2}$, $ S^3$-bundle 
over $S^2 \simeq \mathbb{CP}^1$,
as follows. Let us consider the case $B=\mathbb{CP}^1$.
For the real numbers $\nu_1$ and $\nu_2$ we put 
$\Lambda=4(1-\nu_1^2 \nu_2^2)/(2-\nu_1^2-\nu_2^2)$ , and take the following function 
$X \equiv X_1$ \eqref{EC},
\begin{align}
X(x) = \frac{(x^2-\nu_1^2)(x^2-\nu_2^2)(1-\Lambda x^2/4)}{x^2(x^2-1)} ~.
\end{align}
If we choose the parameters $\{\nu_\alpha \}_{\alpha=1,2}$ as 
\begin{align}\label{intcond} 
k_1 = \frac{\nu_1(1-\nu_2^2)(2-\nu_2^2-\nu_1^2 \nu_2^2)}
      {1+\nu_1^4 \nu_2^2+\nu_1^2 \nu_2^4-3 \nu_1^2 \nu_2^2} ~,~~~
k_2 = \frac{\nu_2(1-\nu_1^2)(2-\nu_1^2-\nu_1^2 \nu_2^2)}
      {1+\nu_1^4 \nu_2^2+\nu_1^2 \nu_2^4-3 \nu_1^2 \nu_2^2} ~,
\end{align}
the corresponding metric \eqref{gCKY} is just the Einstein metric 
found in the paper\cite{Hashimoto:2005} (see Figure \ref{fig:moduli}). 
The case $(k_1,k_2)=(1,1)$ corresponds to the homogeneous Sasaki-Einstein metric 
known in the physics literature as $T^{1,1}$.

%%%%%%%%%%%%%%%%%%%%%%%%%%%%%%%%%%%%%%%%%%%%%%%%%%%
\section{Summary}
%%%%%%%%%%%%%%%%%%%%%%%%%%%%%%%%%%%%%%%%%%%%%%%%%%%
We have reviewed recent developments about exact solutions 
of higher-dimensional Einstein equations
and their symmetries.
Guided by symmetries of the Kerr black hole we introduced conformal Killing-Yano (CKY) tensors.
We forcused mainly on the rank-$2$ closed CKY tensor, which generates mutually commuting Killing tensors
and Killing vectors. The existence of the commuting tensors
underpins the separation of variables in Hamiton-Jacobi, Klein-Gordon and Dirac equations.
The main results are summarized in theorems 5.1-5.3, 
which give a classification of higher-dimensional spacetimes with
a CKY tensor :
\begin{itemize}
\item[$\bullet$] Kerr-NUT-(A)dS black hole spacetime 
is the only Einstein space admitting a principal CKY tensor.
\item[$\bullet$] The most general metrics admitting a rank-$2$ closed 
CKY tensor become Kaluza-Klein metrics \eqref{gCKY} on the bundle over K\"ahler manifolds whose fibers 
are Kerr-NUT-(A)dS spacetimes.
\item[$\bullet$] When the Einstein condition is imposed, 
the metric functions are fixed as \eqref{EC} 
with the K\"ahler-Einstein (and/or general Einstein) base metrics.
\end{itemize}
%This gives a classification of spacetimes admitting a rank-$2$ CKY tensor.\\

Based on these results we further developed the study of Killing-Yano symmetry 
in the presence of skew-symmetric torsion
and presented exact solutions to supergravity theories including the Kerr-Sen black hole. 
Although we did not discuss in this paper,
Dirac operators with skew-symmetric torsion naturally appear 
in the spinorial field equations of supergravity theories\cite{Strominger:1986,Agricola:2006}.
This provides an interesting link to
K\"ahler with torsion (KT) and hyper K\"aler 
with torsion (HKT) manifolds\cite{Howe:1996}\cite{Grantcharov:2000}, 
which have applications across mathematics and physics.

Recently, by Semmelmann\cite{Semmelmann:2002} global properties of CKY tensors were investigated. He showed the existence of CKY tensors on Sasakian, 3-Sasakian,
nearly K\"ahler and weak G$_2$-manifold. These geometries are deeply related to supersymmetric compactifications and AdS/CFT correspondence in string theories\cite{Grana:2006,Acharya:2004}. 
In section 6.2 we presented an explicit method constructing Sasakian manifolds from the generalized Kerr-NUT-(A)dS metrics.
It is an interesting question whether the presented method or its generalizations can provide a new construction in the remaining geometries.

Regarding separability of gravitational perturbation equations,
it is known that the separation of variables occurs 
for some modes in five dimensions \cite{Murata:2008} 
and for tensor modes in higher dimensions \cite{Kunduri:2006b,Oota:2010}.
However, the connection with Killing-Yano symmetry is not clear even four dimensions.
It is important to clarify why the separability works well and also important
to study whether Killing-Yano symmetry enables the separation for more general perturbations.

\section*{Acknowledgements}
We would like to thank V.\ P.\ Frolov, G.\ W.\ Gibbons, P.\ Krtous, D.\ Kubiz\v{n}\'ak, D.\ N.\ Page 
and C.\ M.\ Warnick for the useful discussion. 
We also would like to thank DAMTP, University of Cambridge, for the hospitality.
The work of Y.Y. is supported by the Grant-in Aid for Scientific Research 
No. 21244003 from Japan Ministry of Education.
The work of T.H. is supported by the JSPS Institutional Program for
Young Researcher Overseas Visits
``Promoting international young researchers in mathematics and
mathematical sciences led by OCAMI''. 
T.H. is also grateful to Theoretical Physics Institute, University of Alberta 
and Institute of Theoretical Physics, Charles University for the hospitality.

%\appendix
%\section{First Appendix} %Empty argument \section{} yields `Appendix'. 
%
%\section{Second Appendix}

\end{document}